\documentclass{aa}  
\DeclareUnicodeCharacter{2212}{\textminus}

\usepackage{graphicx}
\usepackage{txfonts}
\usepackage{subcaption}
\usepackage{xcolor}
\usepackage{soul}

\begin{document} 

   \title{The role of inner disk edges in shaping ultra-short-period planet systems around late M dwarfs}

   \subtitle{}

        \author{S. N. Brandenberger,
          \inst{1}          
          M. Sanchez,\inst{1,2}
          N. Van der Marel,\inst{1}
           A. A. Vidotto,\inst{1} 
          Y. Miguel
          \inst{1,3}         
          }

    \institute{Leiden Observatory, Leiden University, P.O. Box 9513, 2300 RA Leiden,
The Netherlands\\
           \email{brandenberger@strw.leidenuniv.nl}
         \and
             Max Planck Institute for Solar System Research, Justus-von-Liebig-Weg 3, 37077 Göttingen, Germany
        \and
            SRON Netherlands Institute for Space Research, Niels Bohrweg 4, 2333 CA Leiden, the Netherlands}

   \date{}

  \abstract  
   {Close-in rocky planets are the most common type of exoplanets around late M dwarfs, ranging from more temperate worlds to highly irradiated lava planets with molten surfaces, and many theoretical studies have attempted to explain their formation. However, the origin of rocky planets with orbital periods shorter than one day, known as ultra-short-period (USP) planets, remains uncertain.}
   {We aim to investigate whether the formation and survival of USP planets is connected to the location of the inner edge of the protoplanetary disk, considering different disk edge prescriptions.}
   {We used N-body simulations that include planet-disk interactions, star-planet tidal interactions, and relativistic corrections, applied to a sample of lunar-mass planetary seeds growing via pebble accretion in a low-viscosity disk ($\alpha_{\rm{t}} = 10^{-4}$). The inner edge of the disk was modeled in three ways: as a fixed close-in edge, as an outward-evolving edge set by the magnetospheric truncation radius, and as an inward-evolving edge defined by the corotation radius.}
   {Ultra-short-period planet formation appears to be tightly controlled by the location of the disk’s inner edge. Our simulations show that only the close-in-fixed-edge scenario and the inward-evolving-edge scenario are capable of producing USP planets, as planets tend to follow the movement of the disk’s inner edge. This suggests that USP planet formation is favored when the inner edge remains close to the corotation radius of a rapidly rotating star.
   }
   {}


   \maketitle

\section{Introduction}

Close-in rocky exoplanets are especially common around M dwarfs \citep[e.g.,][]{DressingandCharbonneau2015, Muldersetal2015, Sabottaetal2021}. Within this population, a non-negligible subset occupies ultra-short-period (USP) orbits, with orbital periods of $P\,<1\,$ day. These planets are estimated to orbit roughly $\sim$0.5\% of G dwarfs, $\sim$0.8\% of K dwarfs, and $\sim$1.1\% of M dwarfs, indicating a mild increase in frequency toward lower-mass hosts \citep{Winnetal2018, Uzsoyetal2021, Hiranoetal2021}. Ultra-short-period planets are likely rocky (with radii $R_p<2R_\oplus$; \citealt{Sanchis-Ojedaetal2014}) and inhabit intensely irradiated and tidal environments \citep{Legeretal2009, Winnetal2018}. Many USPs approach the “lava-world” regime, whereby the day sides of short-period rocky planets reach silicate-melting conditions, potentially forming global magma oceans. This scenario was first proposed for CoRoT-7b and has since been explored for other systems \citep{Legeretal2011, Legeretal2009, Winnetal2017, Zilinskasetal2022}.

Mass inferences for USPs point to small rocky bodies: using \textit{Kepler} data and an Earth-like composition prior, \citet{Uzsoyetal2021} find that most USPs should have $M\sim0.4$\,--\,$8M_\oplus$. Because their sample is dominated by FGK hosts and M-dwarf disks are typically less massive, USPs around M dwarfs may on average possess even lower masses \citep{manara2023demographicsyoungstarsprotoplanetary, vandermarel2024dustevolutionprotoplanetarydisks}.

Recent demographic analyses suggest that the planets below the $P\sim$1 day boundary form a dynamically distinct population. Their radii are systematically smaller than those at $1<P<5$ days, and their period ratios to the nearest companion increase sharply for $P\lesssim2$ days \citep{GoyalandWang2025}. Ultra-short-period planets rarely occur in isolation: in multi-planetary systems, they typically have at least one companion within $P<50$ days \citep{Sanchis-Ojedaetal2014, Adamsetal2021}. These trends point to specialized formation and migration pathways operating at the shortest periods.

Short-period (SP) and USP planets have been modeled with a variety of migration and mass-loss scenarios, with most population studies implementing a fixed inner disk edge, and focusing on solar-type hosts \citep[e.g.,][]{Cossouetal2014,LeeandChiang2017, Izidoroetal2017}. Around M dwarfs, recent modeling adopts fixed inner cavities as well \citep[e.g.,][]{Colemanetal2019, Miguel2020, Ataieeetal2021, Sanchez2024}. 
Although such models reproduce the overall excess of close-in super-Earths, they generally underproduce USPs.
Other studies have allowed the inner edge to evolve in time, usually tying it to the magnetospheric radius so that the cavity moves outward over the disk lifetime.
For Sun-like hosts, it was shown that such outward-moving edges capture and carry resonant chains with the receding disk, and in the giant-planet regime, the subsequent gas dispersal can trigger instabilities within those chains \citep[e.g.,][]{Liuetal2017, Liu2022}.
Recent modeling of the TRAPPIST-1 system also employs an outward-moving inner edge and emphasizes resonance trapping and release as the primary mechanism shaping the system \citep{Pichierrietal2024}. However, these works focus on resonance dynamics and have not yet established whether inner-edge evolution can robustly produce USP systems, especially around late M dwarfs (stars with $M_\star \lesssim 0.35 M_\odot$). 

Building on this indication that close-in planet architectures are likely regulated by the disk’s inner edge, and given the lack of a clear connection between inner-edge physics and USP formation,  
we explore this phenomenon explicitly with this new study where the inner disk edge is time-dependent and test
how its location and evolution shape USP and SP rocky systems around late M dwarfs. We developed N-body simulations that include planet-disk interactions, star-planet tidal interactions, and general relativistic corrections, applied to a sample of lunar-mass embryos that grow by pebble accretion in a low-viscosity disk set by $\alpha=10^{-4}$, following \citet{Sanchez2024}. Our key novelty is the incorporation of new routines that implement distinct, physically motivated prescriptions for the inner-edge evolution. We compare three scenarios: a static, very close-in edge; a magnetospheric-truncation edge that expands outward over the disk lifetime; and a corotation-locked edge that drifts inward as the star spins up. We then assess which prescription reproduces observed features the best, thereby isolating the inner-disk physics required to form USPs around M dwarfs.

This paper is organized as follows. Section \ref{sec:methods} details the numerical framework, with a description of the $N$-body code, the gas and dust disk characterization, and the three inner-edge models (FIX, OUT[M], IN[C]). Section \ref{sec:results} presents the simulation outcomes, from dynamical evolution to encounter and collision statistics, and the resulting planetary system architectures. Section \ref{sec:observations} presents the comparison between our simulated systems and the observed USP population, including the planetary mass-period distribution, the detectability of the simulated planets with TESS and CARMENES, and adjacent period ratios. Section \ref{sec:discussion} puts the results in context by contrasting fixed versus evolving inner-edge models, noting caveats (tides, magnetic torques, and detection biases), and outlining implications for USP demographics and a possible lava-world overlap. Section \ref{sec:conclusions} rounds up the paper with the final conclusions.

\section{Methods}
\label{sec:methods}

This section describes the planet formation model used to simulate the formation and evolution of close-in planetary systems. It summarizes our numerical setup: the $N$-body integrator and its added physics, the host star characterization, the gas and dust disk model, and the simulation setup.

\subsection{Physical processes}

We simulated the formation of close-in planetary systems around a late M dwarf of 0.1 M$_\odot$ with a modified version of the $N$-body code MERCURY \citep{Chambers1999}. We used the identical modifications that add early evolutionary processes relevant to close-in planets as in \citet{Sanchez2024}:
\begin{itemize}
    \item Planet-disk interactions. During the gas-disk phase, we included additional, nongravitational accelerations arising from the torques exerted by the gas disk on embedded embryos and planets. These disk-driven forces modify the planetary dynamics by driving Type-I orbital migration and eccentricity and inclination damping. We adapted the torque prescriptions for low-mass planets from \cite{Paardekooperetal2010, Paardekooperetal2011} and the dynamical friction formulation from \cite{Ida2020} to account for acceleration corrections of planets embedded in the disk. These were applied within a time-evolving gas- disk model based on \cite{Idaetal2016}, which includes two heating mechanisms: viscous heating in the inner disk and irradiation heating in the outer disk. We assumed a smooth surface-density taper at the inner gas-disk edge and computed the standard (two-sided) Type-I torque expressions there using this tapered surface-density profile. In general, the Type-I torque drives inward migration. Nevertheless, close to the inner disk edge, the modified surface-density gradient can alter the torque balance, allowing very low-mass planets (below 0.5 M$_\oplus$) to undergo outward migration (see \cite{Sanchez2022} and \cite{Sanchez2024} for details);
    \item Star-planet tidal interactions. We included acceleration corrections considering both the tides raised by the planet on the star and the tides raised by the star on the planet. These interactions lead to precession of the argument of periastron, secular evolution of the semimajor axis, and eccentricity damping. We modeled tides by using the equilibrium tide formalism of \cite{Hut1981}, adapted to M dwarfs and low-mass planets following \cite{Bolmontetal2011}. Stellar contraction and rotational evolution were treated following \cite{Bolmontetal2011}, using the stellar evolutionary tracks of \cite{Baraffe2015} (for further details, refer to \citealt{Sanchez2020});
    \item General relativistic corrections. We accounted for accelerations acting on the embryos and planets due to general relativistic effects, which induce precession of the argument of periastron. These were implemented via post-Newtonian corrections following \cite{Anderson1975}. (see details in \citealt{Sanchez2020});
    \item Pebble accretion. We assumed that planets grow through pebble accretion following the framework of \cite{LambrechtsAndJohansen2014}, including pebble efficiency prescriptions for noncircular orbits from \cite{LiuAndOrmel2018} and \cite{ OrmelAndLiu2018}. The model distinguishes between 2D and 3D accretion regimes and halts accretion once embryos reach their pebble isolation mass (roughly between 0.5 and 1 M$_\oplus$ within 0.1 au), computed as in \cite{Bitschetal2018}.
\end{itemize}
\subsection{Host star characterization}
\label{sec:host_star_char}
We assumed a 0.1 M$_\odot$ host star and accounted for its pre-main-sequence contraction and the associated spin-up over the course of the integration. The resulting time evolution of the stellar radius and the rotational period is included both in the star-planet tidal interaction calculations and in one of the scenarios explored in this work to evolve the inner edge of the disk.

We adopted the stellar radius evolution $R_\star(t)$ from the models of \cite{Baraffe2015} and followed the stellar rotation period evolution model of \cite{Bolmontetal2011}. In that work, the rotational evolution of very low-mass stars was modeled primarily as contraction-driven spin-up, with an additional term accounting for the influence of orbiting planets. The model was calibrated against observationally inferred rotation rates at different ages from \cite{Herbst2007}. The early spin-up of very low-mass stars during the first few hundred million years \citep[see, e.g.,][]{Bolmont2012} is dominated by stellar contraction. For simplicity, we neglected the rocky planetary contribution, as also assumed in \cite{Bolmontetal2011}. The resulting evolution of the stellar rotation period was mainly determined by the conservation of angular momentum and strongly dependends on the initial stellar rotation period, as follows:
\begin{equation}
P_\star(t) =
P_{\star, \text{ini}}\,
\left(\frac{r_{\mathrm{gyr}}(t)}{r_{\mathrm{gyr}}(t_0)}\right)^{2} 
\left(
\frac{R_\star(t)}{R_0}
\right)^{2}
,\end{equation}
where $P_{\star, \text{ini}}$ denotes the initial rotational period of the star, $R_0$ is the initial stellar radius, and $r_{\textrm{gyr}}$ is the radius of gyration, defined as $r_{\textrm{gyr}}^{2}=I/M_\star R_\star^{2}$, with I the moment of inertia of the host star \citep{Hut1981}. Therefore, assuming spherical symmetry for the host star and a fixed structure constant, the evolution of the stellar rotation period becomes directly linked to the evolution of the stellar radius, as follows:
\begin{equation}
P_\star(t) =
P_{\star, \text{ini}}\,
\left(
\frac{R_\star(t)}{R_0}
\right)^{2}
\label{eq:Prot}
.\end{equation}
We used this relation to compute the stellar rotation period at each timestep of the simulations. This allowed us to consistently track the rotational evolution of the host star throughout the disk lifetime.

\subsection{Disk model}
The dust disk model employed in this study follows the compact, drift-dominated disk model for M dwarfs from \cite{Sanchez2024}, motivated by 
the evolutionary scenario of compact disks proposed by \citet{VanderMarelAndMulders2021} and ALMA observations of compact disks around K and M dwarfs showing typical dust radii of only 2-5 au \citep{Guerra2025}.
The disk was initialized at 1 Myr with a 2 au outer dust radius and an initial pebble mass of $\sim$10 M$_\oplus$, delivered from an initially extended disk of $\sim$20 au inward at a constant pebble mass flux of $5\times10^{-5}$ M$_\oplus$ yr$^{-1}$ (as in \citealt{Sanchez2024}). For the turbulence parameter, $\alpha_t = 10^{-4}$ was adopted, consistent with the low mid-plane viscosity expected in such disks \citep{Rosotti2023,Taboneetal2025}.
Pebble sizes, scale heights, and compositions follow the prescription of \cite{LambrechtsAndJohansen2014} with icy pebbles outside the snowline and dry pebbles within. Under this prescription, the particle Stokes number is typically $St\sim0.02-0.05$.

For the gas disk model, the two-zone model of \cite{Idaetal2016} was adopted as implemented in \cite{Sanchez2024}, with an inner viscous region and an outer irradiation-dominated region. Profiles for $\Sigma_g$, $T_g$, and $h_g$ were computed using the stellar luminosity from \cite{Baraffe2015} and a time-evolving accretion rate based on the empirical fit by \cite{Manara2012}. We adopted a viscosity parameter, $\alpha_g=10^{-4}$, which results in an initial gas disk mass of approximately $10\%$ of the host star mass. We evolved the disk for 10 Myr, consistent with disk lifetimes inferred for low-mass stars \citep{Bayoetal2012,Downesetal2015,Pfalzner2022}. For simplicity, we assumed that the disk is fully dispersed at this time. While we do not explicitly model the physical removal of gas, processes such as photoevaporation and magnetically driven winds are expected to clear the remaining gas on comparable timescales \citep[e.g.,][]{Alexander2014}.
This approach follows common practice in planet formation studies \citep[e.g.,][]{LambrechtsAndJohansen2014,Bitsch2015,Sanchez2024}.

We note that we assumed a smooth surface-density decay at the inner gas-disk edge as in \citep{Sanchez2022,Sanchez2024}:
\begin{equation}
\Sigma_{g,\mathrm{vis}}^{\mathrm{taper}}(r)
= \Sigma_{g,\mathrm{vis}}^{0}(r)\,
\tanh\!\left(\frac{r-r_{\mathrm{inner}}}{r_{\mathrm{inner}}\,h_{\mathrm{inner}}}\right)
,\end{equation}
where $r_{\mathrm{inner}}$ and $h_{\mathrm{inner}}$ are the radius and aspect ratio at the inner edge, respectively, and $\Sigma_{g,\mathrm{vis}}$ is the viscous gas surface density profile. The resulting disk structure sets the torque and damping timescales that govern migration, as well as eccentricity and inclination evolution in the simulations.

In this work new routines were implemented in the N-body code to account for an evolving inner edge of the gas-disk. We explore both static and time-dependent inner edge prescriptions. These are expected to directly impact the planet-disk interactions and pebble accretion processes and are described in the following section.

\begin{figure*}[h!]
\centering
\includegraphics[width=0.96\textwidth]{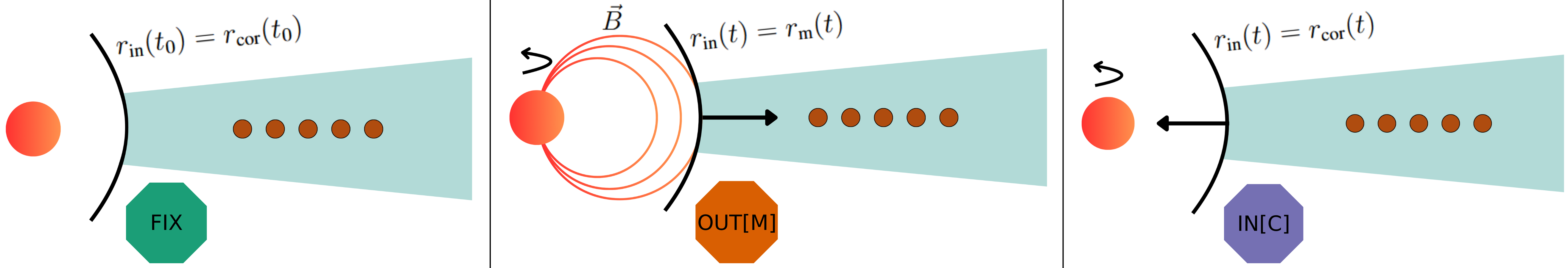}
\caption{Inner disk edge scenarios cartoons. From left to right: FIX Scenario with close-in fixed inner disk edge; OUT[M] Scenario with outward-moving edge based on magnetic flux conservation; IN[C] Scenario with inward-moving edge based on stellar corotation.}
\label{fig:cartoons}
\end{figure*}

\begin{figure}[h!]
\centering
\includegraphics[width=240pt]{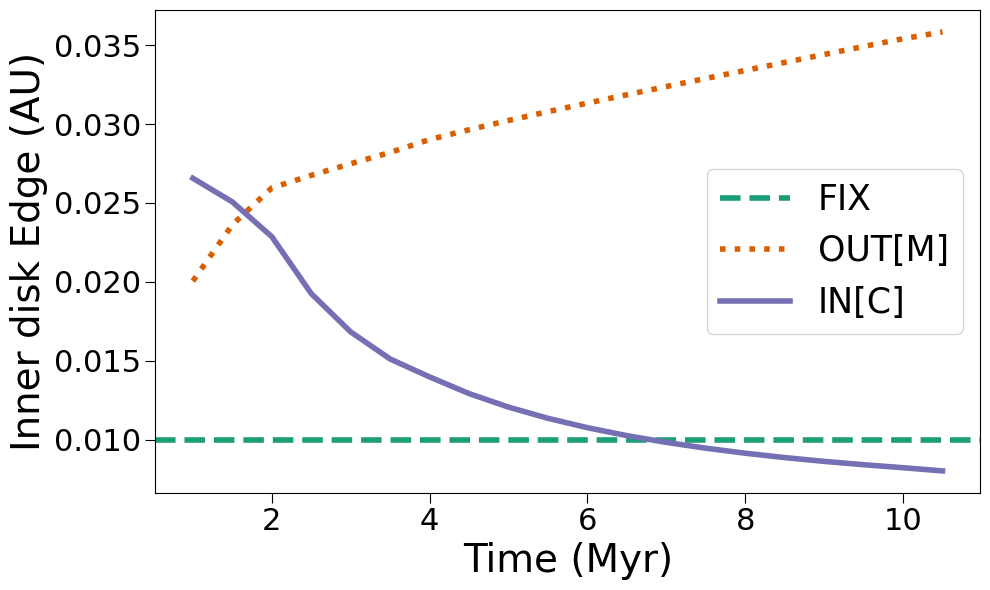}
\caption{Inner disk edge evolution. Fixed inner edge (FIX; dashed green line), outward-migrating edge (OUT[M]; dotted red line; see Eq. \ref{eq:rin_description}), and inward-migrating edge (IN[C]; solid purple line; see Eq. \ref{eq:corotation_rin}). These \(r_{\rm in}(t)\) tracks set where Type-I migration halts.}
\label{fig:Innerdiskedgeevolv}
\end{figure}

\subsection{Inner gas-disk edge models}
\label{sec:diskscenarios}

We explore three prescriptions that differ only in the movement and the physical driver of the inner edge of the disk, $r_{\rm in}$, as shown schematically in Fig. \ref{fig:cartoons}. They are labeled accordingly:
FIX = constant $r_{\rm in}$, OUT = $r_{\rm in}$ increases with time, IN = $r_{\rm in}$ decreases with time. 
One-letter tags indicate the driver: \,[M] = magnetospheric truncation, \,[C] = corotation. 

\subsubsection{Scenario OUT[M]: Outward-drifting inner edge}
Here, the inner disk edge, $r_{\rm in}$, is set by the magnetospheric truncation radius, which is located where the stellar magnetosphere disrupts the accreting gas flow \citep[e.g.,][]{Frank1992}.
Following the normalized approach by \cite{Miguel2020}, we calculated it via
\begin{equation}
\label{eq:rin_description}
\begin{aligned}
r_{\text{in}}(t) &= r_{\text{m}}(t) = 0.01 \left( \frac{B_{\star}(t)}{180\,\mathrm{G}} \right)^{\tfrac{4}{7}}
\left( \frac{R_{\star}(t)}{0.5\,R_{\odot}} \right)^{\tfrac{12}{7}}
\left( \frac{M_{\star}}{0.1\,M_{\odot}} \right)^{-\tfrac{1}{7}} \\
&\times \left( \frac{\dot{M}_\star(t)}{10^{-10}\,M_{\odot}\,\mathrm{yr}^{-1}} \right)^{-\tfrac{2}{7}} \, \text{au},
\end{aligned}
\end{equation}
where $R_{\star}(t)$ is the evolving stellar radius, based on evolutionary tracks by \cite{Baraffe2015}, $M_{\star}$ is the stellar mass, and $\dot{M}_{\star}$ the stellar mass accretion rate based on the empirical fit by \cite{Manara2012}: 
\begin{equation}
\begin{split}
\log\!\left(\frac{\dot{M}_\star(t)}{M_\odot\,\mathrm{yr}^{-1}}\right)
&= -5.12 - 0.46\,\log\!\left(\frac{t}{\mathrm{yr}}\right)
- 5.75\,\log\!\left(\frac{M_\star}{M_\odot}\right) \\
&\quad + 1.17\,\log\!\left(\frac{t}{\mathrm{yr}}\right)
\log\!\left(\frac{M_\star}{M_\odot}\right).
\end{split}
\end{equation}
In this scenario, we let $r_{\rm in}$ evolve by assuming magnetic-flux conservation as the star contracts \cite[e.g.,][]{Reiners2009}. The surface dipole is then
\[
B_\star(t)=B_0\,\left(\frac{R_0}{R_\star(t)}\right)^2,\quad B_0=200\,\mathrm{G},\ R_0=1.003\,R_\odot.
\]
We note that this approximation is valid only over short timescales, and in our model specifically during the disk lifetime ($\sim$10 Myr). Over this period, the stellar radius decreases to about 0.4 $R_\odot$ and the magnetic field strength approaches $\sim$ 1 kG, which is consistent with measured early-age values for very low-mass stars \citep{Morin2008, Donati2008}. 
Over the disk lifetime, $r_{\rm in}$ increases from $\simeq0.02$ au to $\simeq0.036$ au. 
The outward evolution of $r_{\rm in}$ is primarily driven by the decline of $\dot{M}_\star(t)$ through the $\dot{M}_\star^{-2/7}$ dependence in Eq. (\ref{eq:rin_description}). Nevertheless, the magnetic-flux conservation assumed in this study allows the stellar magnetic field to increase over time, contributing roughly a factor of two to the outward evolution. We want to stress that assuming a fixed field would not produce the noticeable outward migration observed. Finally, for comparison, around $t\sim1$ Myr, $r_{\rm in}$ lies close to the corotation radius $r_{\rm cor}$ for $P_\star\approx3$ days (see the following section).

\subsubsection{Scenario IN[C]: Inward-drifting inner edge}
In this prescription, we tied the inner disk edge directly to the evolving stellar corotation radius, which is defined as the distance from the star at which the orbital period equals the stellar rotation period, i.e.,
\begin{equation}
\label{eq:corotation_rin}
r_{\rm in }(t) = r_{\rm cor} (t) =\left(\frac{G M_\star P_\star(t)^2}{4\pi^2}\right)^{1/3},
\end{equation}
where $P_{\star, \text{ini}}=5$\,days and $P_\star(t)$ follows Eq. \eqref{eq:Prot} (see Sect. \ref{sec:host_star_char} for details).
As the star contracts during the pre-main-sequence phase, its rotation period decreases and the corotation radius migrates inward accordingly. Across the disk lifetime, the inner edge therefore moves from 
$\simeq0.027$\,au to $\simeq0.008$\,au, representing continuous disk-locking to the evolving stellar spin. 
\\\\
\begin{table*}[h!]
    \centering
    \caption{Summary of scenarios explored in this study.}
    \label{tab:diskscenarios}
    \begin{tabular*}{\textwidth}{@{\extracolsep{\fill}}p{2.6cm}p{8.2cm}p{2.2cm}}
        \hline
        Scenario & Prescription for $r_{\rm in}$ & Range of $r_{\rm in}$ [au] \\ \hline

        FIX & Fixed inner edge at $P_\star = 1$ d (near corotation). & constant at 0.01 \\ \hline

        OUT[M] & Magnetospheric truncation with $B_\star \propto R_\star^{-2}$ (initial $B_0=200$ G, $R_0=1.003\,R_\odot$); $r_{\rm in} = r_\text{m}(t)$ drifts outward with time. & 0.020--0.036 \\ \hline

        IN[C]  & $r_{\rm in}=r_{\rm cor}(t)$ with $P_{\star, \text{ini}} = 5$ d; $r_{\rm in}$ contracts $\rightarrow$ moves inward with stellar spin-up. & 0.027--0.008 \\ \hline
    \end{tabular*}
    \tablefoot{Rotation values correspond to the stellar period for which $r_{\rm in} \approx r_{\rm cor}$ at 1 Myr. The resulting range of the inner disk edge during the disk lifetime (up to $\sim$10 Myr) is also listed.}
\end{table*}
\subsubsection{Scenario FIX: Fixed inner edge}
In our simplest case, we fixed the inner disk edge at $r_{\rm in} = 0.01$ au for the entire disk lifetime. This fixed edge can be interpreted in two ways: if tied to corotation, it corresponds to $P_\star = 1$ day for a $0.1 M_\odot$ star (see Eq. \ref{eq:corotation_rin}), i.e., right at the USP limit. 
Alternatively, if interpreted as the magnetospheric truncation radius, this location would imply a very low field of $B_{\star} = 60$ G (see Eq. \ref{eq:rin_description}). Thus, this scenario mimics the common “fixed edge” assumption in formation studies, though at a closer-in radius than typically adopted. 

Table \ref{tab:diskscenarios} summarizes the assumptions and resulting $r_{\rm in}$ ranges for each case.
Figure \ref{fig:Innerdiskedgeevolv} shows the temporal evolution of $r_{\rm in}$ for the three scenarios explored in this work.

Note that this study extends the work of \cite{Sanchez2024}, who adopted a fixed inner disk edge of $\approx 0.02$ au and found that no USP planets could be formed with this inner disk edge location in their scenarios. To enable a direct comparison with their results, scenarios OUT[M] and IN[C] were therefore initialized with a truncation radius of $\approx 0.02$ au. Scenario FIX instead adopts a smaller inner edge to 
probe more extreme truncation conditions that may favor USP formation, whereas Scenario IN[C] 
represents an intermediate configuration among the evolving-edge cases, starting slightly 
farther out than OUT[M] and ending slightly closer-in than the fixed configuration. 
We highlight that the initial stellar periods assumed in these scenarios are consistent with the observed range for young low-mass stars \citep[e.g.,][]{IrwinAndBouvier2009, Newtonetal2016}.
Together, these three prescriptions span a wide range of plausible inner disk edge evolutions, from fixed to inward- or outward-drifting, enabling us to test how each pathway affects the formation of USP and SP planets.

\subsection{Code setup}

We conducted 30 N-body simulations, corresponding to ten integrations per scenario. All simulations use the same integration settings, stellar properties, and initial embryo distributions to enable a direct comparison across scenarios.

\subsubsection{Code characterization}

To resolve both long-term evolution and close encounters efficiently, we adopted MERCURY's "hybrid symplectic/Bulirsch-Stoer integrator" for this study. Collisions between embryos were treated as perfect mergers.
We adopted 50 Myr as our nominal integration time to capture late dynamical evolution after disk dispersal (at 10 Myr). A subset of simulations was extended up to 100 Myr to test the impact of longer-term evolution, while a small number were stopped earlier, once their architectures had stabilized (typically after $\gtrsim 30$ Myr). These variations in integration time do not affect our main statistical conclusions.
An ejection threshold of 100 au was applied and the integration counts a close encounter whenever the separation between any pair of bodies falls below three mutual Hill radius. To sufficiently resolve the orbits of USP planets, we chose a time step of 0.04 days, which corresponds to approximately 1/10 of the shortest orbital periods observed around late M dwarfs
(\footnote{https://exoplanetarchive.ipac.caltech.edu/}{Nasa Exoplanet Archive}; \citealt{Christiansenetal2025}).
The stellar radius to account for collisions with the star was assumed to be $\sim$0.4 R$_\odot$, which is an average value for the stellar radius during integration time. 

\subsubsection{Initial embryo distribution}
Each simulation starts with 25 planetary embryos of lunar mass (0.01 M$_\oplus$), 
distributed between 0.5 and 2 au, i.e., exterior to the snowline at 
$r_{\text{snow}} = 0.28$ au. The choice of lunar-mass embryos follows the standard setup used in planet formation studies, also around M dwarfs (e.g., \citealt{KokuboAndIda1998}; \citealt{Liuetal2019}; \citealt{Sanchez2024}). Embryos are initially separated by 15 mutual Hill radii and shared identical bulk properties with a density of 5 g cm$^{-3}$. 

We initialized each embryo with a small eccentricity ($e < 0.02$) and inclination 
($i < 0.5^\circ$), drawing $e$ and $i$ independently from uniform distributions within these ranges. The angular orbital elements, such as argument of periastron ($\omega$), longitude 
of ascending node ($\Omega$), and mean anomaly ($M$), were likewise drawn independently from a uniform distribution between $0^\circ$ and $360^\circ$. 
For a given system index, this complete set of orbital elements (for all 25 embryos) is generated once and then reused identically across all disk scenarios. This procedure 
ensures that differences between outcomes can be attributed exclusively to the treatment of the inner disk edge, rather than to stochastic variations in the initial conditions.

\section{Numerical results}
\label{sec:results}

We present the outcome of 30 N-body simulations of planet formation around a late M dwarf of a stellar mass of 0.1 M$_\odot$ with three different inner disk-edge prescriptions. We quantify how the location and time evolution of the inner disk edge shape orbital migration, dynamical activity (close encounters and collisions), and the final system architectures, including the incidence of USP planets. 

\begin{figure*}[h!]
\centering
\includegraphics[width=0.85\textwidth]{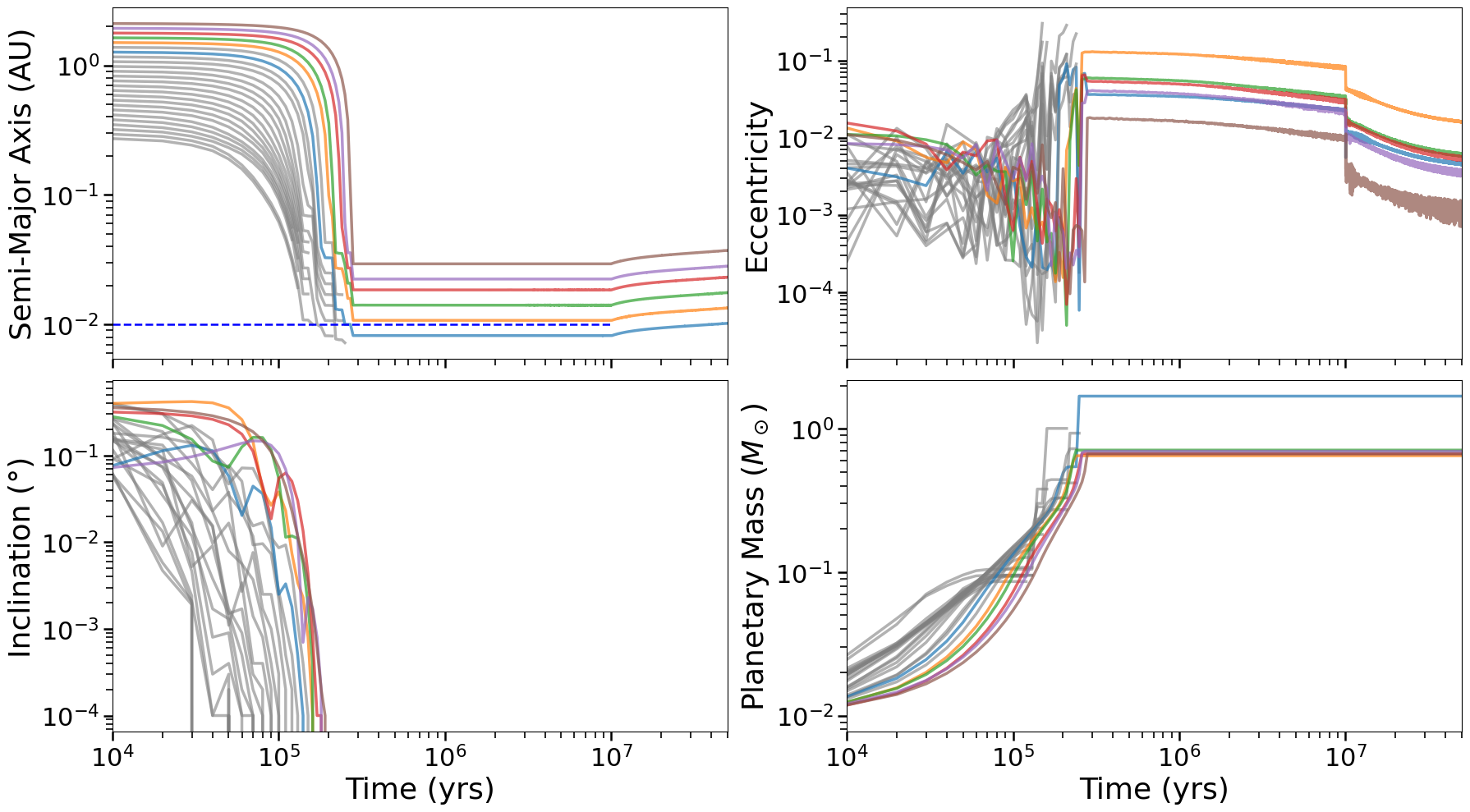}
\caption{Dynamical evolution of planetary embryos in a representative simulation of Scenario FIX. Each line corresponds to one embryo: gray lines show all embryos, while colored lines highlight those that survive until the end of the integration at $\sim$ 50 Myr. Panels display the semimajor axis (top left), eccentricity (top right), inclination (bottom left), and planetary mass (bottom right). The horizontal dashed blue line in the semimajor-axis panel marks the fixed inner disk edge at 0.01 au, which is applied only during the gas-disk phase (up to 10 Myr).}
\label{fig:dynamical_evolution}
\end{figure*}

\subsection{Dynamical evolution}
To illustrate the typical processes that shape our systems, we show in Fig. \ref{fig:dynamical_evolution} the evolution of semimajor axis, eccentricity, inclination, and mass for one representative run of Scenario FIX (one of the two setups that produced USP planets). Overall, this representative system highlights the main evolutionary phases common across the simulations: (i) early convergent migration and collisions in the gas disk, (ii) pile-up and stabilization near the disk edge, and (iii) tidal reshaping of the innermost orbits after disk dispersal. The moving inner edge primarily controls the final semimajor-axis distribution, and thus the USP yield, while the eccentricity and inclination histories remain broadly similar across scenarios.

Across all 30 runs, the qualitative history is similar. Embryos undergo rapid inward Type-I migration and growth by pebble accretion and embryo-embryo collisions within $\sim$$(1$-$3)\times10^{5}$ yr. By this time, most of the mass budget is assembled, and later impacts only perturb masses slightly. Convergent migration drives the embryos to the inner disk edge, where they accumulate and near-commensurate resonant chains form. Typically, a subset of embryos can penetrate into the cavity. Gas damping keeps eccentricities low (typically $e\sim10^{-2}$) and inclinations are rapidly flattened to $\lesssim 0.1°$, leaving nearly coplanar chains.

Once the embryos are trapped in these near-commensurate chains, migration stalls as embryos accumulate at the cavity boundary. In the evolving-edge cases (OUT[M] and IN[C]), the moving truncation or corotation radius shears the chain, briefly breaking and re-capturing resonances. This repacking triggers close encounters and occasional late collisions between Earth-mass planets, producing stair-step changes in the semimajor axis evolution as planets are redistributed to new resonances. 

The main systematic difference is the final parking radius of the chain relative to the USP boundary, set by the inner-edge kinematics.
Scenario FIX stalls planets at a constant radius, where usually many planets manage to penetrate into the cavity in the gas-rich phase. As the inner disk edge is located just at the USP boundary, the innermost planets therefore typically end up interior to the 1 day period line inside the cavity.
In Scenario OUT[M], the magnetospheric truncation radius expands with time, shifting the chain outward. This means that systems that momentarily approach the USP regime early are carried back to longer periods.
In Scenario IN[C], the corotation radius recedes inward, dragging the chain deeper into the cavity and within the USP boundary, efficiently producing USPs in the gas-rich phase.

Once the gas is gone (by $\sim$10 Myr), occasional late collisions between embryos occur and tidal torques secularly reshape the innermost orbits: planets located outside the stellar corotation radius (which is very close-in at this point at $\sim$0.008 au) experience positive torques and migrate outward, while the planetary eccentricities decrease. This tidal interaction explains why some systems retain USP planets while others shift their innermost planets beyond the one-day orbital period. The outward movement is strongest for the closest-in planets. As a result, several systems that produced USPs during the gas phase subsequently move just beyond the USP threshold by the end of the simulation.  
Over gigayear timescales, subsequent stellar spin-down could move the corotation radius outward again and further modify the innermost architecture (see Sect. \ref{sec:discussion}).

\subsection{Collisional history and close encounters}

\begin{figure}[h!]
\centering
\includegraphics[width=220pt]{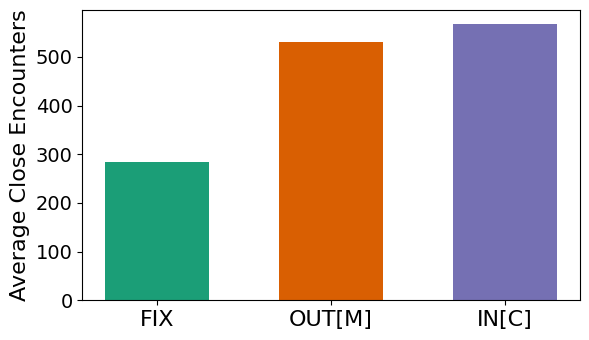}
\caption{Average close-encounter events per system for each inner-edge scenario (FIX, OUT[M], IN[C]). One Hill radius was applied as the threshold. For a given scenario, we sum the number of close-encounter records across all embryos and runs, and divide by the number of valid runs.}
\label{fig:close_encounters}
\end{figure}

\begin{figure}[h!]
\centering
\includegraphics[width=230pt]{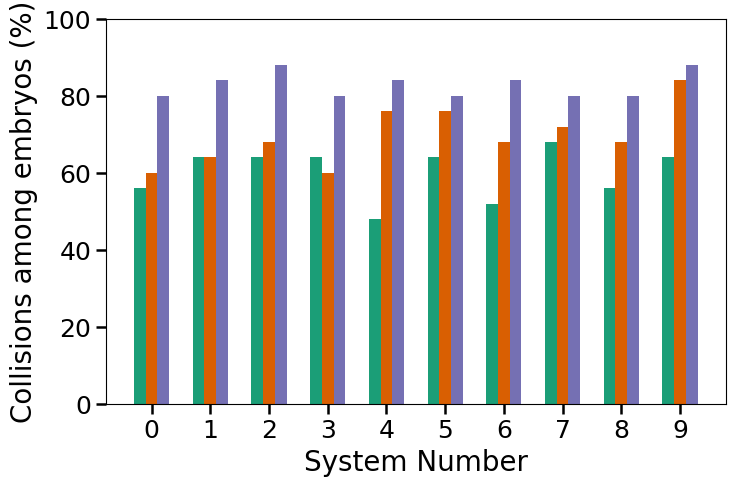}
\caption{Fraction of initial embryos per system that experienced an embryo-embryo collision. Percentages are shown for all 10 systems in each scenario. Bar colors correspond to the three inner-edge prescriptions (green = FIX, orange = OUT[M], purple = IN[C]).}
\label{fig:collisions}
\end{figure}

We evaluated the dynamical activity using four complementary indicators: close-encounter frequency, embryo-embryo mergers, embryo-star collisions, and ejections. Figure \ref{fig:close_encounters} displays the close-encounter statistics, and Fig. \ref{fig:collisions} summarizes embryo-embryo collision outcomes.
We can see that the highest encounter rates occur in Scenario IN[C], which can possibly be attributed to the biggest movement of the disk: from the furthest location to innermost location in our scenarios, due to being linked to 
the stellar spin up. A high encounter rate, however, does not automatically imply stronger instability: in Scenario OUT[M] (second-highest encounter rate), gas damping moderates many interactions so that systems remain dynamically active yet settle into stable configurations.

Embryo-embryo mergers are likewise most frequent in IN[C], consistent with enhanced dynamical mixing as the edge moves inward. OUT[M] shows intermediate merger rates, whereas FIX produces the fewest. 
In contrast, collisions with the central star occur primarily in FIX and IN[C], reflecting their closer inner edges. In FIX, embryos collide with the star in all 10 systems, with an average of 3-4 planet-star collisions per system. In IN[C], collisions with the star occur in about 70\% of runs, with only 1-2 such events per system on average. In OUT[M], the more distant edge tends to delay or prevent such infall. Planetary ejections are rare across all scenarios (two in IN[C], one in FIX, none in OUT[M]), and can be associated with scenarios in which many close interactions occur within resonant systems, which can excite the eccentricity of one of the planets, eventually unbinding it from the system.

\begin{figure}[h!]
\centering
\includegraphics[width=250pt]{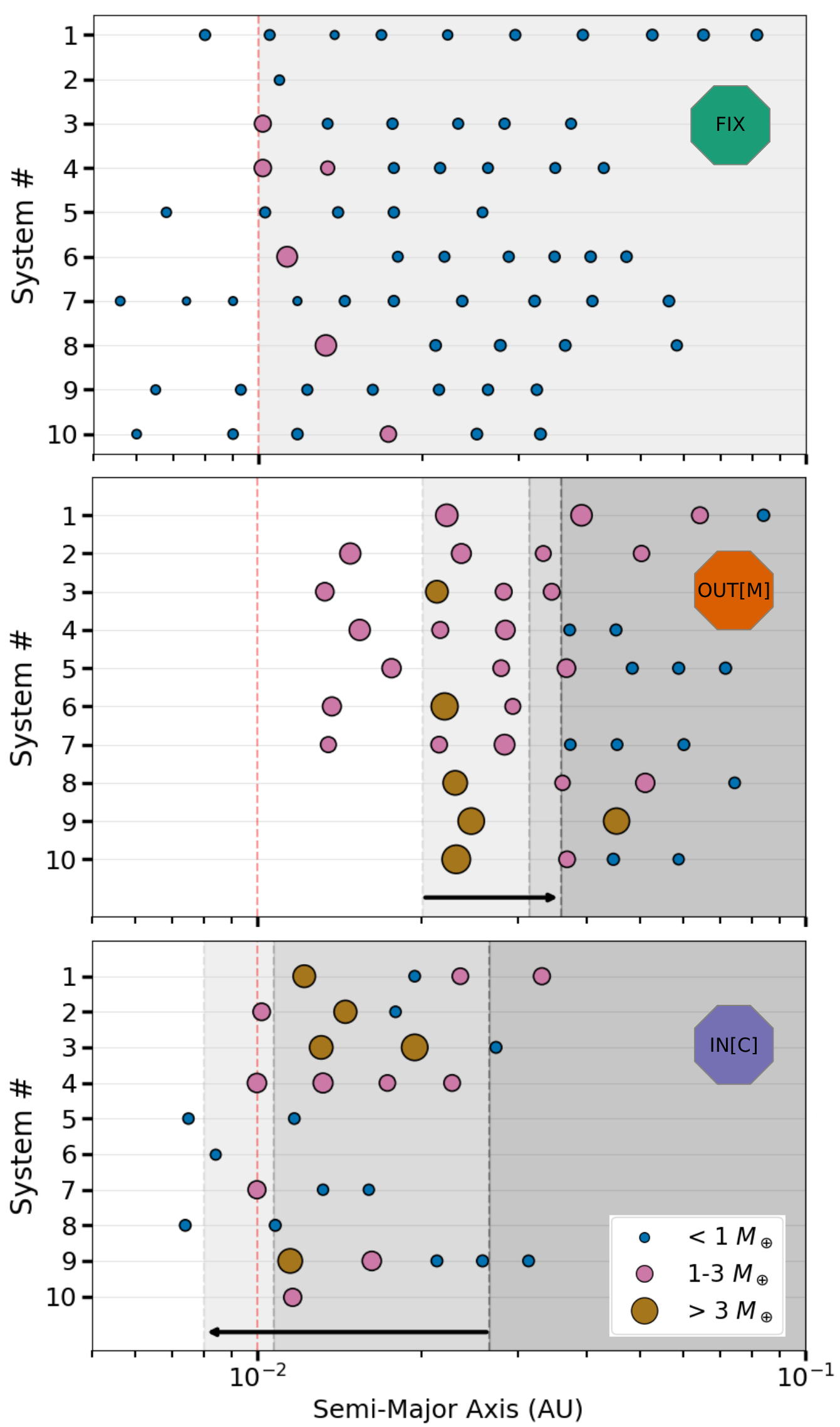}
\caption{Final planetary architectures for the three inner-edge prescriptions at $\sim$ 50 Myr (FIX, OUT[M], and IN[C]). Circles mark planets at the end of each run, where the size and color correspond to their mass: $\sim0.4-1\,M_\oplus$ in blue, $1-3\,M_\oplus$ in pink, and $3-6 \,M_\oplus$ in brown. The dashed red line indicates the approximate 1-day period for a ($0.1\,M_\odot$) star (USP boundary). The shaded band shows the range swept by $r_{\rm in}$ during the gas phase, with dashed gray snapshots at 1, 5, and 10 Myr (or constant) and an arrow for the direction of motion (constant in FIX, outward in OUT[M], inward in IN[C]).}
\label{fig:planetary_architectures}
\end{figure}

\subsection{Planetary architectures}
In Fig. \ref{fig:planetary_architectures} the final semimajor axes and planetary masses are given for each system simulated in each scenario of the study, showing the resulting system architectures.
Despite identical initial conditions, the different prescriptions for the inner disk edge lead to variations in planetary spacing, mass, and multiplicity. 
Only Scenarios FIX and IN[C] form planets interior to the one-day orbital limit, while Scenario OUT[M] yields compact multi-planetary systems without USPs. 

In Scenario FIX, with the inner disk edge fixed at 0.01 au, USP planets consistently form in compact systems of 5-10 planets. 5 out of 10 systems include a USP at the end of the integration. A large majority of the planets is $0.4$-$2.7\,M_\oplus$, and this small range suggests that the mass is more uniformly distributed. The innermost reaches 0.0055 au ($\sim$0.5 d), while outer planets extend to 0.08 au ($\sim$26 d). The most massive planets typically accumulate just outside the disk edge, producing regular orbital spacing and a pronounced mass pile-up near the boundary.

In Scenario OUT[M], where the inner edge expands outward with time, systems host fewer but more massive planets ($0.7$-$5\,M_\oplus$), typically 3-6 on wider orbits up to 0.08 au. No USP planets form, as the innermost orbits are larger than 0.013 au (1.8 d). The outward edge evolution compresses the migration zone, enhances collisions, and yields compact yet less crowded systems that are more massive and diverse than in Scenario FIX.

In Scenario IN[C], with the inner edge contracting inward, USP planets form in 3 out of 10 systems. The systems in this scenario are the least populated, with only 1-5 planets of $0.7$-$4.2\,M_\oplus$, typically between 0.007 and 0.03 au. Showing the lowest multiplicities overall, this scenario produced close-in compact systems that extend less far outward than in Scenario FIX but reach nearly as close to the star.

These architectural differences reflect the underlying dynamical histories. Scenarios with more frequent collisions (OUT[M] and especially IN[C]) yield fewer but more massive planets, whereas the low-collision environment of Scenario FIX preserves higher multiplicities of smaller planets. 

\section{Comparison with observations}
\label{sec:observations}

In this section we test our simulations against observations of M-dwarf systems by analyzing the final planetary masses as a function of orbital period, the detectability of the simulated planets in current surveys, and the spacing between adjacent planets. This allows us to assess whether the models reproduce the observed compact architectures and remain consistent with current detection limits.

\subsection{Planetary mass-orbital period distribution}

We compared the final planetary masses and corresponding orbital periods of the simulated planets across all scenarios explored in this work, FIX, OUT[M], and IN[C], with those of confirmed systems around late M dwarfs ($0.08 < M_\star/M_\odot < 0.35$) hosting USP planets. In total, we identified 13 isolated USP planets and 2 systems containing USP planets with planetary companions, most of which were detected by either \textit{Kepler} or TESS. A few of these systems have radial velocity follow-up observations (e.g., with CARMENES), although the majority lack direct mass measurements. Therefore, for the entire sample, we estimated planetary masses and their uncertainties using the mass-radius power law for rocky planets with radius $R_{\rm p}< 1.6$R$_\oplus$, given by $M_{\rm p} = R_{\rm p}^{3.7}$ \citep{Zeng2016}, and for $R_{\rm p}>1.6$R$_\oplus$, given by $M_{\rm p} = 2.7R_{\rm p}^{1.3}$ \citep{Wolf2016}, with $R_{\rm p}$ in R$_\oplus$ and $M_{\rm p}$ in M$_\oplus$.

\begin{figure}[h!]
\centering
\includegraphics[width=250pt]{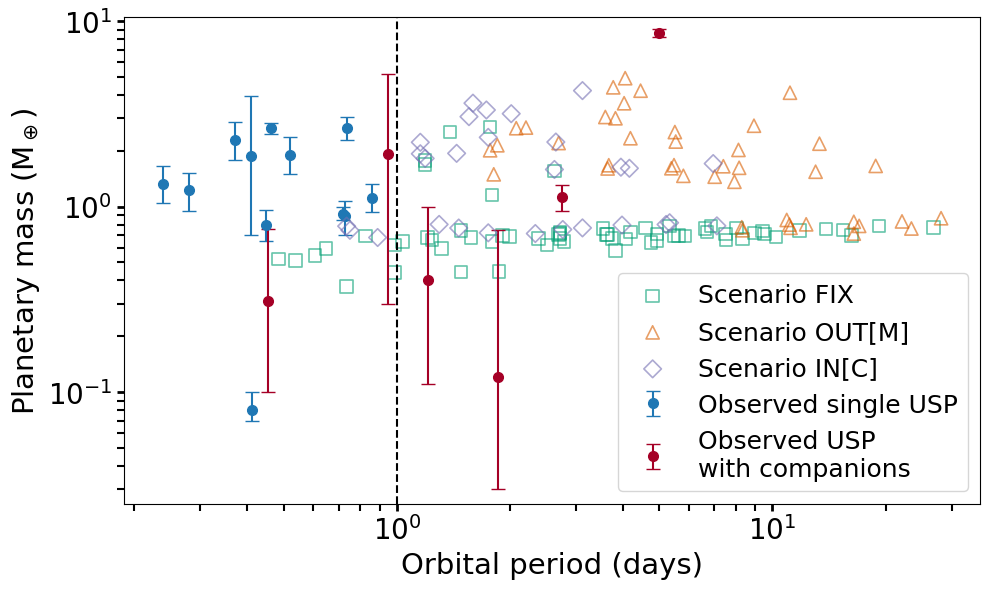}
\caption{Planetary mass as a function of orbital period for our simulated planets in the three inner disk-edge scenarios (FIX, OUT[M], and IN[C]; open symbols) and for the observed USP systems around late M dwarfs ($0.08 < M_\star/M_\odot < 0.35$; filled symbols). Blue circles denote isolated USP planets and red circles USP planets with detected companions. The vertical dashed line marks the 1-day boundary that defines USPs. Asymmetric error bars indicate the $1\sigma$ uncertainties on the observational mass estimates.
}
\label{fig:last_figure}
\end{figure}

In Fig. \ref{fig:last_figure} we present the comparison between our simulated planets and the observed USP sample, including both single and multi-planet systems. Overall, the agreement between simulations and observations is good for each simulation scenario, as we form planets with masses and orbital periods within the observed range. Both the observed systems and our simulations contain only a few planets with inferred masses below 0.5 M$_\oplus$, and for the observed planets these low masses are still associated with large error bars. Our simulations therefore suggest that such low-mass USP planets may exist, but dedicated high-precision radial-velocity follow-up will be required to confirm whether the currently observed candidates are indeed that low in mass. Detecting bodies with masses below 0.5 M$_\oplus$ remains challenging in current observational surveys.
In the following section, we therefore quantify the detection probabilities of our simulated planets with representative surveys.

\begin{figure*}[h!]
\centering
\includegraphics[width=0.99\textwidth]{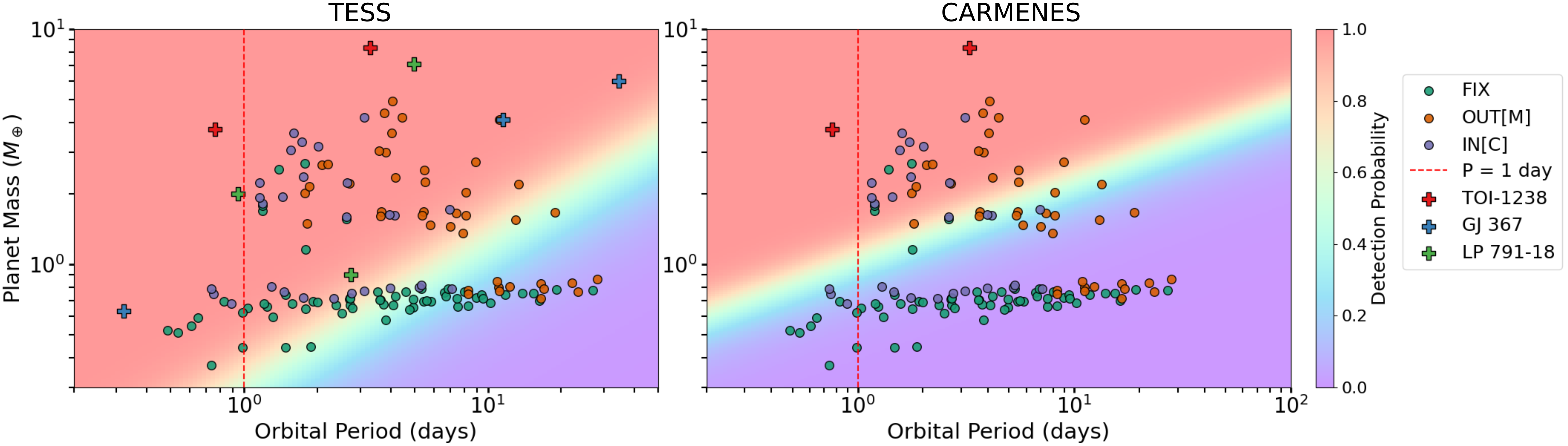}
\caption{Left: Detection probability map for TESS as a function of orbital period and planetary mass. Colors indicate the probability of detecting a transit signal, modeled as a logistic function of S/N with a threshold of 7.3. Known USP systems (TOI-1238, GJ 367, and LP 791-18) detected by TESS are shown for reference.
Right: Detection probability map for CARMENES as a function of orbital period and planetary mass. The model assumes 3~m/s radial velocity precision and 100 observations per target, with a logistic function centered on a S/N of 10. The USP system TOI-1238, which was confirmed by CARMENES, is overplotted. While CARMENES has strong sensitivity to higher-mass planets, the simulated low-mass USP planets fall below the detection threshold, underscoring the challenges of RV confirmation for this population.}
\label{fig:probabilitymaps}
\end{figure*}

\begin{figure}[h!]
\centering
\includegraphics[width=240pt]{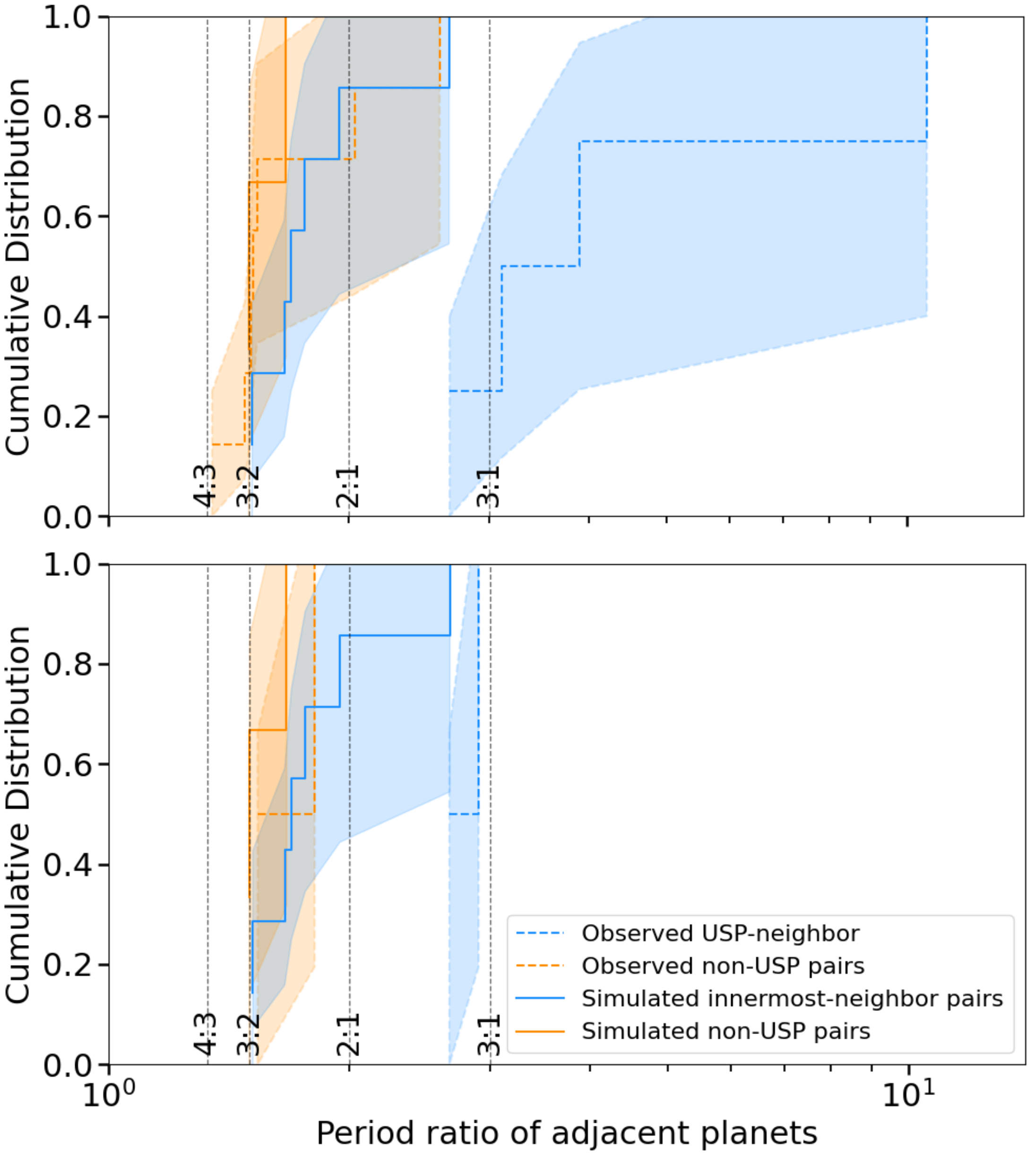}
\caption{Cumulative distributions of adjacent period ratios. Simulated systems in Scenario IN[C] at 50 Myr (solid lines) with detection probabilities associated with TESS higher than 0.9 (see Fig. \ref{fig:probabilitymaps}). Observed exoplanetary systems around stars with masses $0.08<M_\star/M_\odot<0.6$ (top panel) and around stars with masses $0.08<M_\star/M_\odot<0.35$ (bottom panel) that host a USP planet (dashed lines). Observed USP-neighbor pairs are compared to simulated USP + candidate neighbors (blue lines). 
Other adjacent pairs around M dwarfs are compared to all other adjacent pairs in our simulations (orange lines).
Vertical dotted lines mark common MMRs: 4:3, 3:2, 2:1, and 3:1. Poissonian errors for each distribution are shown as shadow areas. The comparison tests whether the early-epoch orbital spacing of prospective USPs resembles the present-day spacing observed for mature USP systems.}
\label{fig:CDF_period_ratios}
\end{figure}

\begin{figure*}[h!]
\centering
\includegraphics[width=0.8\textwidth]{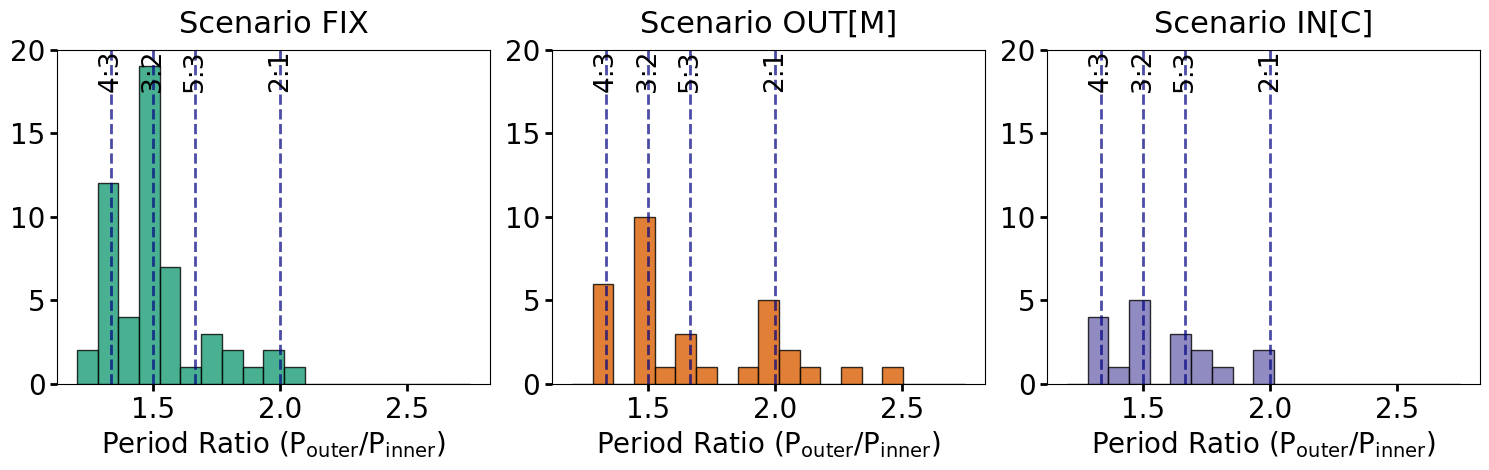}
\caption{Period-ratio distribution of adjacent planet pairs, $P_\text{outer}/P_\text{inner}$, at the end of the integrations for the three inner-edge prescriptions (left to right: FIX, OUT[M], IN[C]). The histograms include all systems formed in each scenario. Vertical dashed lines mark common MMRs (4:3, 3:2, 5:3, 2:1).}
\label{fig:hist_period_ratios}
\end{figure*}

\subsection{Detection probabilities of simulated planets}

To assess which of the simulated planets could be identified with current observational surveys, we constructed detection probability maps for TESS and CARMENES and performed a consistency check with a \textit{Kepler}-based detection model. The maps are based on simplified signal-to-noise ratio (S/N) formulations and serve as first-order estimates of survey sensitivities. They are shown in Fig. \ref{fig:probabilitymaps}.

For TESS, we modeled detectability with a logistic function of the transit S/N, using 7.3 as the threshold value and a steepness parameter of 1.0, following the criteria applied in the mission’s detection pipelines \citep[e.g.,][]{Christiansenetal2015, Sullivanetal2015}. A cadence of 2 minutes and a representative photometric noise level of 2000 ppm were assumed for nearby mid-to-late M dwarfs \citep[values based on][]{Sullivanetal2015, Barclayetal2018}. 
In figure \ref{fig:probabilitymaps}, we present the resulting TESS detection map in the left panel. It shows that planets more massive than about 0.3M$_\oplus$ at ultra-short periods would likely be within TESS’ sensitivity range. Known USP systems observed by TESS are overplotted for comparison \citep{Gonzalezetal2022, Goffoetal2023, Petersonetal2023}, with missing masses calculated like above. \textit{Kepler} probabilities were computed with the same procedure as TESS, using \textit{Kepler} long-cadence noise and duty cycle \citep{DressingandCharbonneau2015, Christiansenetal2015}. However, the results are not shown as they are qualitatively similar. 

For radial velocity detections, we modeled CARMENES's sensitivity by computing the expected RV semi-amplitude $K$ for circular, edge-on orbits and applying a smooth detection probability function (hyperbolic tangent) centered on a conservative S/N = 10. Instrumental performance was set to 3 m/s precision and 100 observations per target, representative of recent survey strategies \citep[e.g.,][]{Gonzalezetal2022, Ribasetal2023}. The CARMENES map, shown in the right panel of Fig. \ref{fig:probabilitymaps}, demonstrates that the instrument is primarily sensitive to higher-mass planets. The low-mass USP planets produced in our simulations fall below its detection threshold. A known USP system confirmed with CARMENES, TOI-1238, is shown for reference \citep{Gonzalezetal2022}.

Overall, the comparison highlights the detection biases of current surveys. We see that TESS is generally sensitive to our simulated USP planets, whereas CARMENES is currently less likely to recover the low-mass short-period planets that dominate our synthetic populations. The known observed USP systems predominantly align with the high-probability regions of the detection maps, lending confidence to the robustness of our modeling approach.
Having established which simulated planets are likely to be detectable with current surveys, we now turn to the orbital architectures of the systems they inhabit.

\subsection{Period ratios of planet pairs}

The spacing of adjacent planets, expressed as orbital period ratios, provides a sensitive diagnostic of system architecture, migration history, and resonance structure.
\textit{Kepler}-wide studies (dominated by FGK hosts) find that USP planets are typically dynamically isolated: their nearest detected neighbor often lies at wide spacings of $P_2/P_1\gtrsim 3$ \citep{SteffenandFarr2013}, which is on average larger than the $\sim$1.3--4 period-ratio range seen among other adjacent planet pairs in \textit{Kepler} multi-planet systems \citep{Fabryckyetal2014}.

To test whether our models reproduce this behavior, we computed period ratios for all adjacent planet pairs at the end of the simulations. For scenarios that produced USPs (FIX and IN[C]),
we include both, planets that already satisfy P$_\mathrm{inner}<1$ day and those whose innermost periods are slightly longer (“USP candidates”).
These are expected to evolve into the USP regime through potential long-term tidal decay (see Sect. \ref{sec:discussion}). We distinguish this combined USP + candidate set from all other adjacent pairs in the corresponding scenario. For the observational comparison, we retrieved \textit{Kepler} DR25 M-dwarf systems from the NASA Exoplanet Archive (\textit{Kepler} DR25, $T\mathrm{eff}<4000$ K or $M_\star<0.6$M$_\odot$; candidates and confirmed planets). Additionally, we compiled a subsample of confirmed systems hosting USP planets detected by other instruments around late M dwarfs ($T_{\mathrm{eff}} < 3400$ K or $M_\star < 0.35M_\odot$). Note that the FGK-dominated literature distribution and our M-dwarf-only comparison need not coincide, owing likely to host-type selection and small-number statistics. 
Figure \ref{fig:CDF_period_ratios} shows the cumulative period-ratio distributions, including their associated Poissonian errors, for systems in Scenario IN[C] at the end of the integration time ($\sim$50 Myr) that are associated with detection probabilities higher than 0.9 (see TESS probability map in Fig. \ref{fig:probabilitymaps}). We note that Scenario IN[C] has a slightly better agreement with observations than Scenario FIX in terms of the period-ratios of innermost-neighbor pairs. For comparison, we also show the observed sample of M-dwarf systems hosting USP planets, as well as the subsample of such systems around late M dwarfs.

In both cases, we find good agreement between the simulated and observed period-ratio distributions for non-USP pairs. However, the simulated USP neighbors are systematically more tightly packed than those observed around M dwarfs, while they show a closer match with the subsample of systems around late M dwarfs. This may indicate that our models underproduce the degree of dynamical isolation inferred for observed USP planets, or that additional planets remain undetected in the current sample. The observational sample is still small, with only four USP-hosting systems around M dwarfs, and just two around late M dwarfs, and is therefore subject to relatively large uncertainties. Moreover, there appears to be a trend toward more compact system architectures around late M dwarfs, which is more consistent with our model predictions.

Figure \ref{fig:hist_period_ratios} shows period-ratio histograms of all adjacent pairs in all systems of all scenarios. These illustrate that the simulations populate common mean-motion resonances (MMRs), especially 4:3 and 3:2, with some pairs near 2:1 and 5:3. Scenario FIX produces the most compact configurations, while Scenario OUT[M] yields a broader distribution, consistent with extended dynamical activity during the gas phase. In Scenario IN[C], the populations concentrate mostly around 4:3 and 3:2, indicating also typical compact architectures.

Together, these comparisons suggest that while our simulations capture the general orbital spacing of multi-planet M-dwarf systems, they do not reproduce the unusually wide USP-neighbor separations seen in observations. Undetected intermediate planets, late collisions between close-in planets, or long-term star-planet tidal interactions could be responsible for these wider separations. However, because long-term simulations are computationally expensive, exploring such effects is outside the scope of this work (see further details in Sect. \ref{sec:discussion}).

\section{Discussion}
\label{sec:discussion}

In this work, we focus on understanding the mechanisms required at the inner edge of the protoplanetary disk to form USP planets or USP planet candidates during the gas disk phase. We find that USP planets form when the disk’s inner edge is located close to the star, around the one-day orbital period limit. In these cases, close-in rocky planets that migrate rapidly inward and remain near the star in compact, resonant or near-resonant chains, tend to follow the evolution of the disk’s inner edge.

Consequently, we observe the formation of USPs in the scenarios where the disk’s inner edge is fixed close to the star assuming an initially fast stellar rotator (with a 1-day period), and in cases where the edge evolves inward following the corotation radius of a star initially rotating with a 5-day period. In contrast, scenarios in which the inner edge migrates outward do not produce USPs. In these cases, no planets are found within 0.015 au (for stellar magnetic field strengths $\gtrsim 200$ G).
\\
These results highlight that USP formation depends both on the absolute location of the inner disk edge and on its evolutionary direction. A fixed inner edge at 0.02 au fails to produce USPs in an otherwise identical setup \citep[see][]{Sanchez2024}, while outward migration further suppresses formation by shifting the trapping location away from the star during the gas disk phase.
Therefore, our results suggest that USP formation requires that the inner disk edge either starts near the 1-day orbital period or migrates inward to reach this close-in location during the gas-disk phase. In the inward-migration case, planet-planet interactions are enhanced and collisions become more frequent, which can produce more massive USP planets. Disk-edge evolution is thus linked not only to whether USPs form, but also to their final masses.

Additional test runs assuming a stellar rotator with a 3-day period, where the disk’s inner edge corresponds to corotation, place the edge as close as 0.0057 au by the end of the gas disk phase. These configurations produce only 0-1 surviving planets per system due to frequent stellar and mutual collisions inside the cavity. This suggests that an inner edge located too close to the star may over-clear the innermost region, drastically reducing system multiplicity and favoring the formation of isolated USPs. Overall, our results emphasize that both the position and temporal evolution of the disk’s inner edge play key roles in regulating USP formation efficiency and the final planetary multiplicity. Close-in architectures thus preserve a dynamical fingerprint of early disk conditions as shown in previous works \citep[e.g.][]{Izidoroetal2017,Lambrechts_2019,Raymond_2022_main,Sanchez2024}. This interpretation is also supported by recent demographic evidence that the location of the innermost small planet correlates with stellar mass and is consistent with regulation by the pre-main-sequence dust sublimation radius of the inner dust disk \citep{Sunetal2025}.

Previous planet formation models that assume a fixed inner cavity for solar-type stars, or a fixed or outward-evolving cavity for M dwarfs, can reproduce close-in rocky planets but generally fail to form USP planets \citep[e.g.,][]{Colemanetal2019, Miguel2020, Ataieeetal2021, Sanchez2024}. This is consistent with our finding that only very close-in or inward-evolving inner edges yield USPs. In contrast to models with outward-moving magnetospheric cavities that emphasize resonance trapping and release \citep[e.g.,][]{Liuetal2017, Liu2022, Pichierrietal2024}, our runs show that such evolution does not generate USPs around late-M stars. This points to the proximity and kinematics of the inner edge, rather than resonance dynamics alone, as the controlling ingredient for USP formation in low-mass systems.

Although we do not model thermal states, the innermost planets produced in Scenario FIX could plausibly reach lava-world equilibrium temperatures. We can therefore interpret our results as also identifying possible dynamical pathways into the molten regime, noting that the lava-world label is temperature-defined and outside the scope of this study.

Our simulations address a disk-regulated pathway in which the inner-edge location and evolution shape the assembly and early dynamical evolution of close-in rocky systems. However, we note that alternative USP-formation channels exist that can appear largely decoupled from the inner disk edge, such as low-eccentricity tidal migration driven by secular interactions and angular-momentum-deficit (AMD) transfer from outer companions \citep[e.g.,][]{PuAndLai2019}. In such cases, the final USP orbit would be set primarily by long-term post-disk secular dynamics and tides, while the inner-edge configuration mainly determines the initial architecture.

For USP orbits, star-planet tidal interactions are important for damping eccentricities and can drive additional orbital migration. We therefore include both tidal forces and general relativistic corrections, as they contribute to the evolution of the argument of periastron and influence close-in orbital evolution \citep{Sanchez2020}. These effects are particularly relevant at USP separations.
Previous work has shown that incorporating tidal interactions enhances the collisional evolution of the innermost planets near the disk’s inner edge, thereby affecting the final masses and compactness of the USP population \citep{Sanchez2020}. This remains true even while planets are still embedded in the gas disk and undergoing planet-disk interactions \citep{Sanchez2022,Sanchez2024}.
Our simulations further show that, during the gas-disk phase, the location and evolution of the inner disk edge primarily set where planets are trapped and thus largely determine the emerging architecture. In this context, tidal effects act as a secondary contribution that further promotes collisions among the innermost planets.

We neglect magnetic star-planet interactions during our integration time. As shown by \citet{Ahuiretal2021}, the relative importance of tides and magnetic torques evolves with age: during the first $\sim$50-100 Myr, tidal torques dominate because the star is still inflated and tides are strongly enhanced. As the star contracts, tidal torques decrease rapidly and magnetic torques can become comparatively more important at later ages even if the stellar magnetic field decays.
Once the star spins down ($\gtrsim$100 Myr), the combined effect of stellar spin evolution and magnetic torques may further affect long-term orbital migration. The sign of the tidal torque depends on the planet’s location relative to the corotation radius: planets located inside (outside) corotation experience inward (outward) tidal torques \citep{Bolmontetal2012}.

In our simulations, several systems that formed USP planets during the gas-rich phase experience a slight outward migration, reaching orbital periods just above one day by 20-30 Myr as a result of tidal interactions with the star. Extending the integrations to include stellar spin-down and magnetic torques could lead to the formation of additional USPs, particularly in systems where the final semimajor axes are smaller than 0.012 au, since secular changes in semimajor axis over gigayear timescales are expected to be less than 10$\%$ \citep{Ahuiretal2021}.
Moreover, because the direction of migration depends on a planet’s position relative to the corotation radius, some USP planets may migrate inward while outer planets continue migrating outward. This differential migration could increase the separation between USPs and their nearest neighbors, potentially reproducing the characteristic 3:1 period ratios observed in real systems. Additionally, such long-term evolution could induce a new phase of dynamical instability, in which some of the closest planets in our simulations (with semimajor axis less than 0.006 au) might eventually collide with the star. However, due to the high computational cost of simulating gigayear timescales, this post-disk evolutionary stage lies beyond the scope of the present work and will be addressed in future studies.

Our relatively small sample size (10 runs per scenario) and the limited integration time of $\sim$50 Myr, together with the detection biases inherent to current transit and radial velocity (RV) surveys (as discussed in Sect. \ref{sec:observations}), restrict a direct comparison with the $\sim1\%$ occurrence rate of close-in planets inferred from observations around M dwarfs \citep{Winnetal2018, Uzsoyetal2021, Hiranoetal2021}. A larger number of simulations extended over gigayear timescales would be required to establish a more quantitative comparison with these occurrence rates. 
In addition, a more detailed analysis between transit and radial-velocity surveys such as \textit{Kepler}, TESS and CARMENES could help understand the fraction of low-mass, close-in planets that may be missing because they currently remain undetected in RV surveys.
Candidate planets showing significant periodic signals are often dismissed when the same signal appears in stellar activity indicators (Rodríguez-López, C., in prep.), which may bias against the detection of such planets.

\section{Conclusions}
\label{sec:conclusions}
We modeled the formation of close-in rocky planets around 0.1 M$_\odot$ stars using a modified $N$-body code that includes the effects of migration, damping, and tidal interactions, starting from compact dust and gas disks extending to 20 au. With the new implementation of evolving inner disk edges in our simulations, we investigated how the resulting truncation shapes the architecture of the innermost planets, focusing on differences between three inner-edge prescriptions. Overall, we find that:
\begin{itemize}
  \item Close-in planets get dynamically coupled to the evolving inner disk edge, migrating with it and assembling into resonant or near-resonant chains.
  \item Ultra-short-period planets only form if the inner edge already lies close to, or sweeps inward across, the USP regime during the gas-rich phase.
  \item A magnetically truncated, outward-moving inner edge (OUT[M]) produces no USPs for the adopted initial stellar magnetic field strength of 200 G.
  \item A corotation-locked inner edge (IN[C]) that sweeps inward can produce USPs with relatively low final multiplicities, whereas a fixed close-in edge (FIX) yields typically lower-mass USPs and higher multiplicities.
  \item Faster initial stellar rotation enhances the USP yield in the corotation-locked scenario, by placing the disk edge initially closer to the star. However, if the edge is too close, the inner cavity is over-cleared.
  \item Stars with initially strong magnetic fields ($\gtrsim 200 G$) are less likely to form USP planets.
\end{itemize}
To conclude, USP formation likely demands a close-in, corotation-controlled inner disk edge during the gas phase. This assumes a fast rotator with a period of $\leq 5$ days.

\begin{acknowledgements}
      This research has made use of the NASA Exoplanet Archive, which is operated by the California Institute of Technology, under contract with the National Aeronautics and Space Administration under the Exoplanet Exploration Program. Furthermore, this work was performed using the compute resources from the Academic Leiden Interdisciplinary Cluster Environment (ALICE) provided by Leiden University.\\
      A.A.V. acknowledges funding from the Dutch Research Council (NWO), with project number VI.C.232.041 of the Talent Programme Vici.\\
      Y.M. acknowledges support from the European Research Council (ERC) under the European Union’s Horizon 2020 research and innovation programme (grant agreement no. 101088557, N-GINE).
\end{acknowledgements}

\bibliographystyle{aa} 
\bibliography{bib} 

\begin{thebibliography}{78}
\expandafter\ifx\csname natexlab\endcsname\relax\def\natexlab#1{#1}\fi

\bibitem[{{Adams} {et~al.}(2021){Adams}, {Jackson}, {Johnson}, {Ciardi},
  {Cochran}, {Endl}, {Everett}, {Furlan}, {Howell}, {Jayanthi}, {MacQueen},
  {Matson}, {Partyka-Worley}, {Schlieder}, {Scott}, {Stanton}, \&
  {Ziegler}}]{Adamsetal2021}
{Adams}, E.~R., {Jackson}, B., {Johnson}, S., {et~al.} 2021, psj, 2, 152

\bibitem[{{Ahuir} {et~al.}(2021){Ahuir}, {Strugarek}, {Brun}, \&
  {Mathis}}]{Ahuiretal2021}
{Ahuir}, J., {Strugarek}, A., {Brun}, A.~S., \& {Mathis}, S. 2021, \aap, 650,
  A126

\bibitem[{{Alexander} {et~al.}(2014){Alexander}, {Pascucci}, {Andrews},
  {Armitage}, \& {Cieza}}]{Alexander2014}
{Alexander}, R., {Pascucci}, I., {Andrews}, S., {Armitage}, P., \& {Cieza}, L.
  2014, in Protostars and Planets VI, ed. H.~{Beuther}, R.~S. {Klessen}, C.~P.
  {Dullemond}, \& T.~{Henning}, 475--496

\bibitem[{{Anderson} {et~al.}(1975){Anderson}, {Esposito}, {Martin},
  {Thornton}, \& {Muhleman}}]{Anderson1975}
{Anderson}, J.~D., {Esposito}, P.~B., {Martin}, W., {Thornton}, C.~L., \&
  {Muhleman}, D.~O. 1975, \apj, 200, 221

\bibitem[{{Ataiee} \& {Kley}(2021)}]{Ataieeetal2021}
{Ataiee}, S. \& {Kley}, W. 2021, aap, 648, A69

\bibitem[{{Baraffe} {et~al.}(2015){Baraffe}, {Homeier}, {Allard}, \&
  {Chabrier}}]{Baraffe2015}
{Baraffe}, I., {Homeier}, D., {Allard}, F., \& {Chabrier}, G. 2015, \aap, 577,
  A42

\bibitem[{{Barclay} {et~al.}(2018){Barclay}, {Pepper}, \&
  {Quintana}}]{Barclayetal2018}
{Barclay}, T., {Pepper}, J., \& {Quintana}, E.~V. 2018, \apjs, 239, 2

\bibitem[{{Bayo} {et~al.}(2012){Bayo}, {Barrado}, {Hu{\'e}lamo},
  {Morales-Calder{\'o}n}, {Melo}, {Stauffer}, \& {Stelzer}}]{Bayoetal2012}
{Bayo}, A., {Barrado}, D., {Hu{\'e}lamo}, N., {et~al.} 2012, \aap, 547, A80

\bibitem[{{Bitsch} {et~al.}(2015){Bitsch}, {Lambrechts}, \&
  {Johansen}}]{Bitsch2015}
{Bitsch}, B., {Lambrechts}, M., \& {Johansen}, A. 2015, \aap, 582, A112

\bibitem[{{Bitsch} {et~al.}(2018){Bitsch}, {Morbidelli}, {Johansen}, {Lega},
  {Lambrechts}, \& {Crida}}]{Bitschetal2018}
{Bitsch}, B., {Morbidelli}, A., {Johansen}, A., {et~al.} 2018, \aap, 612, A30

\bibitem[{{Bolmont} {et~al.}(2011){Bolmont}, {Raymond}, \&
  {Leconte}}]{Bolmontetal2011}
{Bolmont}, E., {Raymond}, S.~N., \& {Leconte}, J. 2011, \aap, 535, A94

\bibitem[{{Bolmont} {et~al.}(2012{\natexlab{a}}){Bolmont}, {Raymond},
  {Leconte}, \& {Matt}}]{Bolmont2012}
{Bolmont}, E., {Raymond}, S.~N., {Leconte}, J., \& {Matt}, S.~P.
  2012{\natexlab{a}}, \aap, 544, A124

\bibitem[{{Bolmont} {et~al.}(2012{\natexlab{b}}){Bolmont}, {Raymond},
  {Leconte}, \& {Matt}}]{Bolmontetal2012}
{Bolmont}, E., {Raymond}, S.~N., {Leconte}, J., \& {Matt}, S.~P.
  2012{\natexlab{b}}, \aap, 544, A124

\bibitem[{{Chambers}(1999)}]{Chambers1999}
{Chambers}, J.~E. 1999, MNRAS, 304, 793

\bibitem[{{Christiansen} {et~al.}(2015){Christiansen}, {Clarke}, {Burke},
  {Seader}, {Jenkins}, {Twicken}, {Catanzarite}, {Smith}, {Batalha}, {Haas},
  {Thompson}, {Campbell}, {Sabale}, \& {Kamal Uddin}}]{Christiansenetal2015}
{Christiansen}, J.~L., {Clarke}, B.~D., {Burke}, C.~J., {et~al.} 2015, \apj,
  810, 95

\bibitem[{{Christiansen} {et~al.}(2025){Christiansen}, {McElroy}, {Harbut},
  {Ciardi}, {Crane}, {Good}, {Hardegree-Ullman}, {Kesseli}, {Lund}, {Lynn},
  {Muthiar}, {Nilsson}, {Oluyide}, {Papin}, {Rivera}, {Swain}, {Susemiehl},
  {Tam}, {van Eyken}, \& {Beichman}}]{Christiansenetal2025}
{Christiansen}, J.~L., {McElroy}, D.~L., {Harbut}, M., {et~al.} 2025, psj, 6,
  186

\bibitem[{{Coleman} {et~al.}(2019){Coleman}, {Leleu}, {Alibert}, \&
  {Benz}}]{Colemanetal2019}
{Coleman}, G.~A.~L., {Leleu}, A., {Alibert}, Y., \& {Benz}, W. 2019, aap, 631,
  A7

\bibitem[{{Cossou} {et~al.}(2014){Cossou}, {Raymond}, {Hersant}, \&
  {Pierens}}]{Cossouetal2014}
{Cossou}, C., {Raymond}, S.~N., {Hersant}, F., \& {Pierens}, A. 2014, aap, 569,
  A56

\bibitem[{{Donati} {et~al.}(2008){Donati}, {Morin}, {Petit}, {Delfosse},
  {Forveille}, {Auri{\`e}re}, {Cabanac}, {Dintrans}, {Fares}, {Gastine},
  {Jardine}, {Ligni{\`e}res}, {Paletou}, {Ramirez Velez}, \&
  {Th{\'e}ado}}]{Donati2008}
{Donati}, J.~F., {Morin}, J., {Petit}, P., {et~al.} 2008, MNRAS, 390, 545

\bibitem[{{Downes} {et~al.}(2015){Downes}, {Rom{\'a}n-Z{\'u}{\~n}iga},
  {Ballesteros-Paredes}, {Mateu}, {Brice{\~n}o}, {Hern{\'a}ndez},
  {Petr-Gotzens}, {Calvet}, {Hartmann}, \& {Mauco}}]{Downesetal2015}
{Downes}, J.~J., {Rom{\'a}n-Z{\'u}{\~n}iga}, C., {Ballesteros-Paredes}, J.,
  {et~al.} 2015, MNRAS, 450, 3490

\bibitem[{{Dressing} \& {Charbonneau}(2015)}]{DressingandCharbonneau2015}
{Dressing}, C.~D. \& {Charbonneau}, D. 2015, \apj, 807, 45

\bibitem[{{Fabrycky} {et~al.}(2014){Fabrycky}, {Lissauer}, {Ragozzine}, {Rowe},
  {Steffen}, {Agol}, {Barclay}, {Batalha}, {Borucki}, {Ciardi}, {Ford},
  {Gautier}, {Geary}, {Holman}, {Jenkins}, {Li}, {Morehead}, {Morris},
  {Shporer}, {Smith}, {Still}, \& {Van Cleve}}]{Fabryckyetal2014}
{Fabrycky}, D.~C., {Lissauer}, J.~J., {Ragozzine}, D., {et~al.} 2014, \apj,
  790, 146

\bibitem[{{Frank} {et~al.}(1992){Frank}, {King}, \& {Raine}}]{Frank1992}
{Frank}, J., {King}, A., \& {Raine}, D. 1992, {Accretion power in
  astrophysics.}, Vol.~21

\bibitem[{{Goffo} {et~al.}(2023){Goffo}, {Gandolfi}, {Egger}, {Mustill},
  {Albrecht}, {Hirano}, {Kochukhov}, {Astudillo-Defru}, {Barragan}, {Serrano},
  {Hatzes}, {Alibert}, {Guenther}, {Dai}, {Lam}, {Csizmadia}, {Smith},
  {Fossati}, {Luque}, {Rodler}, {Winther}, {R{\o}rsted}, {Alarcon}, {Bonfils},
  {Cochran}, {Deeg}, {Jenkins}, {Korth}, {Livingston}, {Meech}, {Murgas},
  {Orell-Miquel}, {Osborne}, {Palle}, {Persson}, {Redfield}, {Ricker},
  {Seager}, {Vanderspek}, {Van Eylen}, \& {Winn}}]{Goffoetal2023}
{Goffo}, E., {Gandolfi}, D., {Egger}, J.~A., {et~al.} 2023, apjl, 955, L3

\bibitem[{{Gonz{\'a}lez-{\'A}lvarez} {et~al.}(2022){Gonz{\'a}lez-{\'A}lvarez},
  {Zapatero Osorio}, {Sanz-Forcada}, {Caballero}, {Reffert}, {B{\'e}jar},
  {Hatzes}, {Herrero}, {Jeffers}, {Kemmer}, {L{\'o}pez-Gonz{\'a}lez}, {Luque},
  {Molaverdikhani}, {Morello}, {Nagel}, {Quirrenbach}, {Rodr{\'\i}guez},
  {Rodr{\'\i}guez-L{\'o}pez}, {Schlecker}, {Schweitzer}, {Stock}, {Passegger},
  {Trifonov}, {Amado}, {Baker}, {Boyd}, {Cadieux}, {Charbonneau}, {Collins},
  {Doyon}, {Dreizler}, {Espinoza}, {F{\H{u}}r{\'e}sz}, {Furlan}, {Hesse},
  {Howell}, {Jenkins}, {Kidwell}, {Latham}, {McLeod}, {Montes}, {Morales},
  {O'Dwyer}, {Pall{\'e}}, {Pedraz}, {Reiners}, {Ribas}, {Quinn}, {Schnaible},
  {Seager}, {Skinner}, {Smith}, {Schwarz}, {Shporer}, {Vanderspek}, \&
  {Winn}}]{Gonzalezetal2022}
{Gonz{\'a}lez-{\'A}lvarez}, E., {Zapatero Osorio}, M.~R., {Sanz-Forcada}, J.,
  {et~al.} 2022, \aap, 658, A138

\bibitem[{{Goyal} \& {Wang}(2025)}]{GoyalandWang2025}
{Goyal}, A.~V. \& {Wang}, S. 2025, \aj, 169, 191

\bibitem[{{Guerra-Alvarado} {et~al.}(2025){Guerra-Alvarado}, {van der Marel},
  {Williams}, {Pinilla}, {Mulders}, {Lambrechts}, \& {Sanchez}}]{Guerra2025}
{Guerra-Alvarado}, O.~M., {van der Marel}, N., {Williams}, J.~P., {et~al.}
  2025, aap, 696, A232

\bibitem[{{Herbst} {et~al.}(2007){Herbst}, {Eisl{\"o}ffel}, {Mundt}, \&
  {Scholz}}]{Herbst2007}
{Herbst}, W., {Eisl{\"o}ffel}, J., {Mundt}, R., \& {Scholz}, A. 2007, in
  Protostars and Planets V, ed. B.~{Reipurth}, D.~{Jewitt}, \& K.~{Keil}, 297

\bibitem[{{Hirano} {et~al.}(2021){Hirano}, {Livingston}, {Fukui}, {Narita},
  {Harakawa}, {Ishikawa}, {Miyakawa}, {Kimura}, {Nakayama}, {Fujita}, {Hori},
  {Stassun}, {Bieryla}, {Cadieux}, {Ciardi}, {Collins}, {Ikoma}, {Vanderburg},
  {Barclay}, {Brasseur}, {de Leon}, {Doty}, {Doyon}, {Esparza-Borges},
  {Esquerdo}, {Furlan}, {Gaidos}, {Gonzales}, {Hodapp}, {Howell}, {Isogai},
  {Jacobson}, {Jenkins}, {Jensen}, {Kawauchi}, {Kotani}, {Kudo}, {Kurita},
  {Kurokawa}, {Kusakabe}, {Kuzuhara}, {Lafreni{\`e}re}, {Latham}, {Massey},
  {Mori}, {Murgas}, {Nishikawa}, {Nishiumi}, {Omiya}, {Paegert}, {Palle},
  {Parviainen}, {Quinn}, {Ricker}, {Schwarz}, {Seager}, {Tamura}, {Tenenbaum},
  {Terada}, {Vanderspek}, {Vievard}, {Watanabe}, \& {Winn}}]{Hiranoetal2021}
{Hirano}, T., {Livingston}, J.~H., {Fukui}, A., {et~al.} 2021, \aj, 162, 161

\bibitem[{{Hut}(1981)}]{Hut1981}
{Hut}, P. 1981, \aap, 99, 126

\bibitem[{{Ida} {et~al.}(2016){Ida}, {Guillot}, \& {Morbidelli}}]{Idaetal2016}
{Ida}, S., {Guillot}, T., \& {Morbidelli}, A. 2016, \aap, 591, A72

\bibitem[{{Ida} {et~al.}(2020){Ida}, {Muto}, {Matsumura}, \&
  {Brasser}}]{Ida2020}
{Ida}, S., {Muto}, T., {Matsumura}, S., \& {Brasser}, R. 2020, MNRAS, 494, 5666

\bibitem[{{Irwin} \& {Bouvier}(2009)}]{IrwinAndBouvier2009}
{Irwin}, J. \& {Bouvier}, J. 2009, in IAU Symposium, Vol. 258, The Ages of
  Stars, ed. E.~E. {Mamajek}, D.~R. {Soderblom}, \& R.~F.~G. {Wyse}, 363--374

\bibitem[{{Izidoro} {et~al.}(2017){Izidoro}, {Ogihara}, {Raymond},
  {Morbidelli}, {Pierens}, {Bitsch}, {Cossou}, \& {Hersant}}]{Izidoroetal2017}
{Izidoro}, A., {Ogihara}, M., {Raymond}, S.~N., {et~al.} 2017, MNRAS, 470, 1750

\bibitem[{{Kokubo} \& {Ida}(1998)}]{KokuboAndIda1998}
{Kokubo}, E. \& {Ida}, S. 1998, \icarus, 131, 171

\bibitem[{{Lambrechts} \& {Johansen}(2014)}]{LambrechtsAndJohansen2014}
{Lambrechts}, M. \& {Johansen}, A. 2014, \aap, 572, A107

\bibitem[{{Lambrechts} {et~al.}(2019){Lambrechts}, {Morbidelli}, {Jacobson},
  {Johansen}, {Bitsch}, {Izidoro}, \& {Raymond}}]{Lambrechts_2019}
{Lambrechts}, M., {Morbidelli}, A., {Jacobson}, S.~A., {et~al.} 2019, aap, 627,
  A83

\bibitem[{{Lee} \& {Chiang}(2017)}]{LeeandChiang2017}
{Lee}, E.~J. \& {Chiang}, E. 2017, \apj, 842, 40

\bibitem[{{L{\'e}ger} {et~al.}(2011){L{\'e}ger}, {Grasset}, {Fegley}, {Codron},
  {Albarede}, {Barge}, {Barnes}, {Cance}, {Carpy}, {Catalano}, {Cavarroc},
  {Demangeon}, {Ferraz-Mello}, {Gabor}, {Grie{\ss}meier}, {Leibacher},
  {Libourel}, {Maurin}, {Raymond}, {Rouan}, {Samuel}, {Schaefer}, {Schneider},
  {Schuller}, {Selsis}, \& {Sotin}}]{Legeretal2011}
{L{\'e}ger}, A., {Grasset}, O., {Fegley}, B., {et~al.} 2011, \icarus, 213, 1

\bibitem[{{L{\'e}ger} {et~al.}(2009){L{\'e}ger}, {Rouan}, {Schneider}, {Barge},
  {Fridlund}, {Samuel}, {Ollivier}, {Guenther}, {Deleuil}, {Deeg}, {Auvergne},
  {Alonso}, {Aigrain}, {Alapini}, {Almenara}, {Baglin}, {Barbieri}, {Bruntt},
  {Bord{\'e}}, {Bouchy}, {Cabrera}, {Catala}, {Carone}, {Carpano}, {Csizmadia},
  {Dvorak}, {Erikson}, {Ferraz-Mello}, {Foing}, {Fressin}, {Gandolfi},
  {Gillon}, {Gondoin}, {Grasset}, {Guillot}, {Hatzes}, {H{\'e}brard}, {Jorda},
  {Lammer}, {Llebaria}, {Loeillet}, {Mayor}, {Mazeh}, {Moutou}, {P{\"a}tzold},
  {Pont}, {Queloz}, {Rauer}, {Renner}, {Samadi}, {Shporer}, {Sotin}, {Tingley},
  {Wuchterl}, {Adda}, {Agogu}, {Appourchaux}, {Ballans}, {Baron}, {Beaufort},
  {Bellenger}, {Berlin}, {Bernardi}, {Blouin}, {Baudin}, {Bodin}, {Boisnard},
  {Boit}, {Bonneau}, {Borzeix}, {Briet}, {Buey}, {Butler}, {Cailleau},
  {Cautain}, {Chabaud}, {Chaintreuil}, {Chiavassa}, {Costes}, {Cuna Parrho},
  {de Oliveira Fialho}, {Decaudin}, {Defise}, {Djalal}, {Epstein}, {Exil},
  {Faur{\'e}}, {Fenouillet}, {Gaboriaud}, {Gallic}, {Gamet}, {Gavalda},
  {Grolleau}, {Gruneisen}, {Gueguen}, {Guis}, {Guivarc'h}, {Guterman},
  {Hallouard}, {Hasiba}, {Heuripeau}, {Huntzinger}, {Hustaix}, {Imad},
  {Imbert}, {Johlander}, {Jouret}, {Journoud}, {Karioty}, {Kerjean},
  {Lafaille}, {Lafond}, {Lam-Trong}, {Landiech}, {Lapeyrere}, {Larqu{\'e}},
  {Laudet}, {Lautier}, {Lecann}, {Lefevre}, {Leruyet}, {Levacher}, {Magnan},
  {Mazy}, {Mertens}, {Mesnager}, {Meunier}, {Michel}, {Monjoin}, {Naudet},
  {Nguyen-Kim}, {Orcesi}, {Ottacher}, {Perez}, {Peter}, {Plasson}, {Plesseria},
  {Pontet}, {Pradines}, {Quentin}, {Reynaud}, {Rolland}, {Rollenhagen},
  {Romagnan}, {Russ}, {Schmidt}, {Schwartz}, {Sebbag}, {Sedes}, {Smit},
  {Steller}, {Sunter}, {Surace}, {Tello}, {Tiph{\`e}ne}, {Toulouse}, {Ulmer},
  {Vandermarcq}, {Vergnault}, {Vuillemin}, \& {Zanatta}}]{Legeretal2009}
{L{\'e}ger}, A., {Rouan}, D., {Schneider}, J., {et~al.} 2009, \aap, 506, 287

\bibitem[{{Liu} {et~al.}(2019){Liu}, {Lambrechts}, {Johansen}, \&
  {Liu}}]{Liuetal2019}
{Liu}, B., {Lambrechts}, M., {Johansen}, A., \& {Liu}, F. 2019, \aap, 632, A7

\bibitem[{{Liu} \& {Ormel}(2018)}]{LiuAndOrmel2018}
{Liu}, B. \& {Ormel}, C.~W. 2018, \aap, 615, A138

\bibitem[{{Liu} {et~al.}(2017){Liu}, {Ormel}, \& {Lin}}]{Liuetal2017}
{Liu}, B., {Ormel}, C.~W., \& {Lin}, D. N.~C. 2017, \aap, 601, A15

\bibitem[{{Liu} {et~al.}(2022){Liu}, {Raymond}, \& {Jacobson}}]{Liu2022}
{Liu}, B., {Raymond}, S.~N., \& {Jacobson}, S.~A. 2022, nat, 604, 643

\bibitem[{Manara {et~al.}(2023)Manara, Ansdell, Rosotti, Hughes, Armitage,
  Lodato, \& Williams}]{manara2023demographicsyoungstarsprotoplanetary}
Manara, C.~F., Ansdell, M., Rosotti, G.~P., {et~al.} 2023, Demographics of
  young stars and their protoplanetary disks: lessons learned on disk evolution
  and its connection to planet formation

\bibitem[{{Manara} {et~al.}(2012){Manara}, {Robberto}, {Da Rio}, {Lodato},
  {Hillenbrand}, {Stassun}, \& {Soderblom}}]{Manara2012}
{Manara}, C.~F., {Robberto}, M., {Da Rio}, N., {et~al.} 2012, \apj, 755, 154

\bibitem[{{Miguel} {et~al.}(2020){Miguel}, {Cridland}, {Ormel}, {Fortney}, \&
  {Ida}}]{Miguel2020}
{Miguel}, Y., {Cridland}, A., {Ormel}, C.~W., {Fortney}, J.~J., \& {Ida}, S.
  2020, MNRAS, 491, 1998

\bibitem[{{Morin} {et~al.}(2008){Morin}, {Donati}, {Petit}, {Delfosse},
  {Forveille}, {Albert}, {Auri{\`e}re}, {Cabanac}, {Dintrans}, {Fares},
  {Gastine}, {Jardine}, {Ligni{\`e}res}, {Paletou}, {Ramirez Velez}, \&
  {Th{\'e}ado}}]{Morin2008}
{Morin}, J., {Donati}, J.~F., {Petit}, P., {et~al.} 2008, MNRAS, 390, 567

\bibitem[{{Mulders} {et~al.}(2015){Mulders}, {Ciesla}, {Min}, \&
  {Pascucci}}]{Muldersetal2015}
{Mulders}, G.~D., {Ciesla}, F.~J., {Min}, M., \& {Pascucci}, I. 2015, \apj,
  807, 9

\bibitem[{{Newton} {et~al.}(2016){Newton}, {Irwin}, {Charbonneau},
  {Berta-Thompson}, {Dittmann}, \& {West}}]{Newtonetal2016}
{Newton}, E.~R., {Irwin}, J., {Charbonneau}, D., {et~al.} 2016, \apj, 821, 93

\bibitem[{{Ormel} \& {Liu}(2018)}]{OrmelAndLiu2018}
{Ormel}, C.~W. \& {Liu}, B. 2018, \aap, 615, A178

\bibitem[{{Paardekooper} {et~al.}(2010){Paardekooper}, {Baruteau}, {Crida}, \&
  {Kley}}]{Paardekooperetal2010}
{Paardekooper}, S.~J., {Baruteau}, C., {Crida}, A., \& {Kley}, W. 2010, MNRAS,
  401, 1950

\bibitem[{{Paardekooper} {et~al.}(2011){Paardekooper}, {Baruteau}, \&
  {Kley}}]{Paardekooperetal2011}
{Paardekooper}, S.~J., {Baruteau}, C., \& {Kley}, W. 2011, MNRAS, 410, 293

\bibitem[{{Peterson} {et~al.}(2023){Peterson}, {Benneke}, {Collins}, {Piaulet},
  {Crossfield}, {Ali-Dib}, {Christiansen}, {Gagn{\'e}}, {Faherty}, {Kite},
  {Dressing}, {Charbonneau}, {Murgas}, {Cointepas}, {Almenara}, {Bonfils},
  {Kane}, {Werner}, {Gorjian}, {Roy}, {Shporer}, {Pozuelos}, {Socia},
  {Cloutier}, {Dietrich}, {Irwin}, {Weiss}, {Waalkes}, {Berta-Thomson},
  {Evans}, {Apai}, {Parviainen}, {Pall{\'e}}, {Narita}, {Howard}, {Dragomir},
  {Barkaoui}, {Gillon}, {Jehin}, {Ducrot}, {Benkhaldoun}, {Fukui}, {Mori},
  {Nishiumi}, {Kawauchi}, {Ricker}, {Latham}, {Winn}, {Seager}, {Isaacson},
  {Bixel}, {Gibbs}, {Jenkins}, {Smith}, {Chavez}, {Rackham}, {Henning},
  {Gabor}, {Chen}, {Espinoza}, {Jensen}, {Collins}, {Schwarz}, {Conti}, {Wang},
  {Kielkopf}, {Mao}, {Horne}, {Sefako}, {Quinn}, {Moldovan}, {Fausnaugh},
  {F{\.z}{\.z}r{\'e}sz}, \& {Barclay}}]{Petersonetal2023}
{Peterson}, M.~S., {Benneke}, B., {Collins}, K., {et~al.} 2023, nat, 617, 701

\bibitem[{{Pfalzner} {et~al.}(2022){Pfalzner}, {Dehghani}, \&
  {Michel}}]{Pfalzner2022}
{Pfalzner}, S., {Dehghani}, S., \& {Michel}, A. 2022, apjl, 939, L10

\bibitem[{{Pichierri} {et~al.}(2024){Pichierri}, {Morbidelli}, {Batygin}, \&
  {Brasser}}]{Pichierrietal2024}
{Pichierri}, G., {Morbidelli}, A., {Batygin}, K., \& {Brasser}, R. 2024, Nature
  Astronomy, 8, 1408

\bibitem[{Pu \& Lai(2019)}]{PuAndLai2019}
Pu, B. \& Lai, D. 2019, Monthly Notices of the Royal Astronomical Society, 488,
  3568

\bibitem[{Raymond \& Morbidelli(2022)}]{Raymond_2022_main}
Raymond, S.~N. \& Morbidelli, A. 2022, Planet Formation: Key Mechanisms and
  Global Models (Springer International Publishing), 3--82

\bibitem[{{Reiners} {et~al.}(2009){Reiners}, {Scholz}, {Eisl{\"o}ffel},
  {Hallinan}, {Berger}, {Browning}, {Irwin}, {K{\"u}ker}, \&
  {Matt}}]{Reiners2009}
{Reiners}, A., {Scholz}, A., {Eisl{\"o}ffel}, J., {et~al.} 2009, in American
  Institute of Physics Conference Series, Vol. 1094, 15th Cambridge Workshop on
  Cool Stars, Stellar Systems, and the Sun, ed. E.~{Stempels} (AIP), 250--257

\bibitem[{{Ribas} {et~al.}(2023){Ribas}, {Reiners}, {Zechmeister}, {Caballero},
  {Morales}, {Sabotta}, {Baroch}, {Amado}, {Quirrenbach}, {Abril}, {Aceituno},
  {Anglada-Escud{\'e}}, {Azzaro}, {Barrado}, {B{\'e}jar}, {Ben{\'\i}tez de
  Haro}, {Bergond}, {Bluhm}, {Calvo Ortega}, {Cardona Guill{\'e}n},
  {Chaturvedi}, {Cifuentes}, {Colom{\'e}}, {Cont}, {Cort{\'e}s-Contreras},
  {Czesla}, {D{\'\i}ez-Alonso}, {Dreizler}, {Duque-Arribas}, {Espinoza},
  {Fern{\'a}ndez}, {Fuhrmeister}, {Galad{\'\i}-Enr{\'\i}quez},
  {Garc{\'\i}a-L{\'o}pez}, {Gonz{\'a}lez-{\'A}lvarez}, {Gonz{\'a}lez
  Hern{\'a}ndez}, {Guenther}, {de Guindos}, {Hatzes}, {Henning}, {Herrero},
  {Hintz}, {Huelmo}, {Jeffers}, {Johnson}, {de Juan}, {Kaminski}, {Kemmer},
  {Khaimova}, {Khalafinejad}, {Kossakowski}, {K{\"u}rster}, {Labarga},
  {Lafarga}, {Lalitha}, {Lamp{\'o}n}, {Lillo-Box}, {Lodieu}, {L{\'o}pez
  Gonz{\'a}lez}, {L{\'o}pez-Puertas}, {Luque}, {Mag{\'a}n}, {Mancini},
  {Marfil}, {Mart{\'\i}n}, {Mart{\'\i}n-Ruiz}, {Molaverdikhani}, {Montes},
  {Nagel}, {Nortmann}, {Nowak}, {Pall{\'e}}, {Passegger}, {Pavlov}, {Pedraz},
  {Perdelwitz}, {Perger}, {Ram{\'o}n-Ballesta}, {Reffert}, {Revilla},
  {Rodr{\'\i}guez}, {Rodr{\'\i}guez-L{\'o}pez}, {Sadegi}, {S{\'a}nchez
  Carrasco}, {S{\'a}nchez-L{\'o}pez}, {Sanz-Forcada}, {Sch{\"a}fer},
  {Schlecker}, {Schmitt}, {Sch{\"o}fer}, {Schweitzer}, {Seifert}, {Shan},
  {Skrzypinski}, {Solano}, {Stahl}, {Stangret}, {Stock}, {St{\"u}rmer},
  {Tabernero}, {Tal-Or}, {Trifonov}, {Vanaverbeke}, {Yan}, \& {Zapatero
  Osorio}}]{Ribasetal2023}
{Ribas}, I., {Reiners}, A., {Zechmeister}, M., {et~al.} 2023, \aap, 670, A139

\bibitem[{{Rosotti}(2023)}]{Rosotti2023}
{Rosotti}, G.~P. 2023, nar, 96, 101674

\bibitem[{{Sabotta} {et~al.}(2021){Sabotta}, {Schlecker}, {Chaturvedi},
  {Guenther}, {Mu{\~n}oz Rodr{\'\i}guez}, {Mu{\~n}oz S{\'a}nchez}, {Caballero},
  {Shan}, {Reffert}, {Ribas}, {Reiners}, {Hatzes}, {Amado}, {Klahr}, {Morales},
  {Quirrenbach}, {Henning}, {Dreizler}, {Pall{\'e}}, {Perger}, {Azzaro},
  {Jeffers}, {Kaminski}, {K{\"u}rster}, {Lafarga}, {Montes}, {Passegger}, \&
  {Zechmeister}}]{Sabottaetal2021}
{Sabotta}, S., {Schlecker}, M., {Chaturvedi}, P., {et~al.} 2021, \aap, 653,
  A114

\bibitem[{{Sanchez} {et~al.}(2024){Sanchez}, {van der Marel}, {Lambrechts},
  {Mulders}, \& {Guerra-Alvarado}}]{Sanchez2024}
{Sanchez}, M., {van der Marel}, N., {Lambrechts}, M., {Mulders}, G.~D., \&
  {Guerra-Alvarado}, O.~M. 2024, \aap, 689, A236

\bibitem[{{S{\'a}nchez} {et~al.}(2020){S{\'a}nchez}, {de El{\'\i}a}, \&
  {Downes}}]{Sanchez2020}
{S{\'a}nchez}, M.~B., {de El{\'\i}a}, G.~C., \& {Downes}, J.~J. 2020, \aap,
  637, A78

\bibitem[{{S{\'a}nchez} {et~al.}(2022){S{\'a}nchez}, {de El{\'\i}a}, \&
  {Downes}}]{Sanchez2022}
{S{\'a}nchez}, M.~B., {de El{\'\i}a}, G.~C., \& {Downes}, J.~J. 2022, \aap,
  663, A20

\bibitem[{{Sanchis-Ojeda} {et~al.}(2014){Sanchis-Ojeda}, {Rappaport}, {Winn},
  {Kotson}, {Levine}, \& {El Mellah}}]{Sanchis-Ojedaetal2014}
{Sanchis-Ojeda}, R., {Rappaport}, S., {Winn}, J.~N., {et~al.} 2014, \apj, 787,
  47

\bibitem[{{Steffen} \& {Farr}(2013)}]{SteffenandFarr2013}
{Steffen}, J.~H. \& {Farr}, W.~M. 2013, \apjl, 774, L12

\bibitem[{{Sullivan} {et~al.}(2015){Sullivan}, {Winn}, {Berta-Thompson},
  {Charbonneau}, {Deming}, {Dressing}, {Latham}, {Levine}, {McCullough},
  {Morton}, {Ricker}, {Vanderspek}, \& {Woods}}]{Sullivanetal2015}
{Sullivan}, P.~W., {Winn}, J.~N., {Berta-Thompson}, Z.~K., {et~al.} 2015, \apj,
  809, 77

\bibitem[{{Sun, Meng-Fei} {et~al.}(2025){Sun, Meng-Fei}, {Xie, Ji-Wei}, {Zhou,
  Ji-Lin}, {Liu, Beibei}, {Nikolaou, Nikolaos}, \& {Millholland, Sarah
  C.}}]{Sunetal2025}
{Sun, Meng-Fei}, {Xie, Ji-Wei}, {Zhou, Ji-Lin}, {et~al.} 2025, A\&A, 699, A333

\bibitem[{{Tabone} {et~al.}(2025){Tabone}, {Rosotti}, {Trapman}, {Pinilla},
  {Pascucci}, {Somigliana}, {Alexander}, {Vioque}, {Anania}, {Kuznetsova},
  {Zhang}, {P{\'e}rez}, {Cieza}, {Carpenter}, {Deng}, {Agurto-Gangas},
  {Ruiz-Rodriguez}, {Sierra}, {Kurtovic}, {Miley}, {Gonz{\'a}lez-Ruilova},
  {TorresVillanueva}, {Hogerheijde}, {Schwarz}, {Toci}, {Testi}, \&
  {Lodato}}]{Taboneetal2025}
{Tabone}, B., {Rosotti}, G.~P., {Trapman}, L., {et~al.} 2025, \apj, 989, 7

\bibitem[{{Uzsoy} {et~al.}(2021){Uzsoy}, {Rogers}, \& {Price}}]{Uzsoyetal2021}
{Uzsoy}, A. S.~M., {Rogers}, L.~A., \& {Price}, E.~M. 2021, \apj, 919, 26

\bibitem[{{van der Marel} \& {Mulders}(2021)}]{VanderMarelAndMulders2021}
{van der Marel}, N. \& {Mulders}, G.~D. 2021, \aj, 162, 28

\bibitem[{van~der Marel \&
  Pinilla(2024)}]{vandermarel2024dustevolutionprotoplanetarydisks}
van~der Marel, N. \& Pinilla, P. 2024, Dust evolution in protoplanetary disks

\bibitem[{{Winn} {et~al.}(2018){Winn}, {Sanchis-Ojeda}, \&
  {Rappaport}}]{Winnetal2018}
{Winn}, J.~N., {Sanchis-Ojeda}, R., \& {Rappaport}, S. 2018, \nar, 83, 37

\bibitem[{{Winn} {et~al.}(2017){Winn}, {Sanchis-Ojeda}, {Rogers}, {Petigura},
  {Howard}, {Isaacson}, {Marcy}, {Schlaufman}, {Cargile}, \&
  {Hebb}}]{Winnetal2017}
{Winn}, J.~N., {Sanchis-Ojeda}, R., {Rogers}, L., {et~al.} 2017, \aj, 154, 60

\bibitem[{{Wolfgang} {et~al.}(2016){Wolfgang}, {Rogers}, \& {Ford}}]{Wolf2016}
{Wolfgang}, A., {Rogers}, L.~A., \& {Ford}, E.~B. 2016, apj, 825, 19

\bibitem[{{Zeng} {et~al.}(2016){Zeng}, {Sasselov}, \& {Jacobsen}}]{Zeng2016}
{Zeng}, L., {Sasselov}, D.~D., \& {Jacobsen}, S.~B. 2016, apj, 819, 127

\bibitem[{{Zilinskas} {et~al.}(2022){Zilinskas}, {van Buchem}, {Miguel},
  {Louca}, {Lupu}, {Zieba}, \& {van Westrenen}}]{Zilinskasetal2022}
{Zilinskas}, M., {van Buchem}, C.~P.~A., {Miguel}, Y., {et~al.} 2022, \aap,
  661, A126

\end{thebibliography}

\end{document}